\documentclass[11pt,headings=big,tightenlines,numbers=noenddot,DIV=14,a4paper]{article}%
\pdfoutput=1

\usepackage[utf8]{inputenc}
\usepackage[left=2.55cm, right=2.55cm, top=2.55cm, bottom=2.55cm]{geometry}
\usepackage{amsmath,amssymb,amsbsy,amsfonts,amssymb,amsthm,mathtools}
\usepackage{bm,relsize}  
\usepackage{slashed}
\usepackage[normalem]{ulem}
\usepackage{xcolor}
\usepackage{graphicx}
\usepackage{url}
\usepackage{cancel}
\usepackage{cite}
\usepackage[colorlinks=true,allcolors=darkpurple,pdfborder={0 0 0},linktocpage=false]{hyperref}
\usepackage{tabularx,booktabs}
\usepackage{multicol}
\usepackage{units}
\usepackage{xspace}
\usepackage[labelfont=bf]{caption}
\usepackage[section]{placeins}
\usepackage{subcaption}
\usepackage{soul} % para tachar
\usepackage{changepage}

\definecolor{darkred}{rgb}{0.6,0,0}
\definecolor{darkpurple}{rgb}{0.5,0,0.5}

\def\hc{\text{h.c.}}

\def\IM{\, \text{Im}}
\def\RE{\, \text{Re}}

\def\z2{$\mathbb{Z}_2$}
\def\321{$\mathrm{SU(3)_c} \times \mathrm{SU(2)_L} \times \mathrm{U(1)_Y}$}

\def\cvac			{c_{\text{vac}}}
\def\Tnuc			{T_{\text{nuc}}}
\def\RH			{\mathsmaller{\text{RH}}}
\def\BAU			{\mathsmaller{\text{BAU}}}
\def\TC			{\mathsmaller{\text{TC}}}
\def\SM			{\mathsmaller{\text{SM}}}
\def\GW			{{\mathsmaller{\text{GW}}}}
\def\Pl			{\mathsmaller{\text{Pl}}}
\def\gwp			{\gamma}
\def\gcoll			{\gamma_\text{coll}}
\def\gwpT			{\gamma_{\mathsmaller{\text{T}}}}
\def\gwprun		{\gamma_{\text{run}}}
\def\PT 			{\mathsmaller{\text{PT}}}

\newcommand{\RN}[1]{%
  \textup{\uppercase\expandafter{\romannumeral#1}}%
}

\definecolor{vdrgreen}{rgb}{0.0, 0.7, 0.0}
\definecolor{avblue}{rgb}{0.0, 0.0, 0.8}

\newcommand{\AddrUNIBO}{%
\textit{\small%
Dipartimento di Fisica e Astronomia, Università di Bologna, via Irnerio 46, 40126 Bologna, Italy}}

\newcommand{\AddrINFNBO}{%
\textit{\small INFN, Sezione di Bologna, viale Berti Pichat 6/2, 40127 Bologna, Italy
}}
\newcommand{\AddrLPTHE}{%
\textit{\small 
LPTHE, CNRS \& Sorbonne Université, 4 Place Jussieu, F-75252, Paris, France
}}

\begin{document}

\begin{center}
\vspace*{15mm}

{\huge \bf 
Baryogenesis and Leptogenesis\\ \vspace{.3 cm}
from Supercooled Confinement
} \\
\vspace{1cm}

{
{\large Maximilian Dichtl}$^{\text{a,b}}$, {\large Jacopo Nava}$^{\text{b,c}}$, {\large Silvia Pascoli}$^{\text{b,c}}$, {\large Filippo Sala}$^{\text{b,c}}$\footnote{\textit{On leave of absence from LPTHE, CNRS \& Sorbonne Universit\'{e}, Paris, France.}}}

 \vspace*{.5cm} 
 $^{(\text{a})}$ \AddrLPTHE  \\\vspace*{.2cm} 
 $^{(\text{b})}$\AddrUNIBO \\\vspace*{.2cm} 
 $^{(\text{c})}$ \AddrINFNBO  
 % \\\vspace*{.2cm} $^{(\text{d})}$ \AddrCERN
\end{center}

\vspace*{10mm}
\begin{abstract}\noindent\normalsize
We propose a framework of baryogenesis and leptogenesis that relies on a supercooled confining phase transition (PT) in the early universe. The baryon or lepton asymmetry is sourced by decays of hadrons of the strong dynamics after the PT,
%and it is washed out weakly
and it is enhanced compared to the non-confining case, which was the only one explored so far. This widens the energy range of the PT, where the observed baryon asymmetry %is reproduced
can be reproduced, down to the electroweak scale. The framework then becomes testable with gravity waves (GW) at LISA and the Einstein Telescope. We then study two explicit realisations: one of leptogenesis from composite sterile neutrinos that realises inverse see-saw; one of baryogenesis from composite scalars that is partly testable by existing colliders and flavour factories.
\end{abstract}

\newpage
\par\noindent\rule{\textwidth}{0.5pt}
\setcounter{tocdepth}{2}
\tableofcontents
\par\noindent\rule{\textwidth}{0.5pt}

\vspace*{5mm}
\section{Introduction}

%Observations at the scale of our solar system~\cite{}, all the way up~\cite{} to the scale of the entire universe~\cite{BBN, CMB}, consistently inform us that the universe contains way more matter than antimatter.
The universe contains way more matter than antimatter. Independent measurements of the number density $n_b$ of baryons (i.e. particles with a non-zero baryon number $B$) with the cosmic microwave background~\cite{Planck:2018vyg} and big bang nucleosynthesis~\cite{Fields:2019pfx}, combined with the lack of evidence for primordial antibaryons~\cite{Cohen:1997ac,Steigman:2008ap}, imply a baryon asymmetry of the universe (BAU)
\begin{equation}
Y_\BAU
\equiv (n_b - n_{\bar{b}})/s
\simeq 8.6 \times 10^{-11}\,,
\end{equation}
where $s$ is the entropy density. The Standard Model of particle physics (SM), when combined with the standard cosmological model $\Lambda$CDM, clashes with the observed BAU (see e.g.~\cite{Kolb:1990vq}) unless one imposes it as an ad-hoc initial condition, a possibility which becomes untenable when combined with the inflationary paradigm. The BAU is then a strong observational motivation for physics beyond the SM.

To possibly reproduce the observed BAU, one needs processes in the early universe which, at the same time~\cite{Sakharov:1967dj}: i) violate of course $B$; ii) violate also $C$ and $CP$, otherwise any non-zero $B$ in their final state would be compensated by a $-B$ in the final state of the conjugate process; iii) are out of equilibrium, otherwise the reverse processes would wash-out the baryon asymmetry.
Cosmological first order phase transitions (PTs) satisfy condition iii), so that they can potentially explain the BAU if they are accompanied by processes that violate $B$, $C$ and $CP$.
While first-order PTs in the early universe are not predicted by the SM in isolation~\cite{Aoki:2006we,Kajantie:1996mn}, they are by many SM extensions motivated by open theoretical problems, see e.g.~\cite{Creminelli:2001th,Nardini:2007me,Konstandin:2011dr,Craig:2020jfv,Jinno:2016knw,DelleRose:2019pgi,VonHarling:2019rgb,Greljo:2019xan}.
The proposal of electroweak baryogenesis, for example, relies exactly on new physics that makes the EW PT first-order, see~\cite{Bodeker:2020ghk} for a recent review.

First-order PTs with fast bubble walls, in particular, are currently receiving massive attention because they produce gravitational waves (GW)~\cite{Witten:1984rs,Hogan:1986qda} observable at current~\cite{NANOGrav:2023hvm,Antoniadis:2023zhi,Gouttenoire:2023bqy} and foreseen~\cite{LISACosmologyWorkingGroup:2022jok} telescopes.
% LISA WG ref  also includes sensitivites of AdvLIGO, ET and others
In addition to detectability in GW, first-order PTs with fast walls are motivated as production mechanisms of dark matter (DM) in the form of very heavy particles~\cite{Falkowski:2012fb,Hambye:2018qjv,Baldes:2020kam,Azatov:2021ifm,Baldes:2022oev,Baldes:2023fsp} or primordial black holes~\cite{Liu:2021svg,Hashino:2021qoq,Lewicki:2023ioy,Gouttenoire:2023naa}.
PTs with fast bubble walls offer also viable mechanisms to explain the baryon asymmetry, see e.g.~\cite{Katz:2016adq,Azatov:2021irb,Baldes:2021vyz}. A particularly simple possibility is to obtain the BAU from out-of-equilibrium decays of particles that obtain their mass at a supercooled PT, as proposed in~\cite{Baldes:2021vyz} and worked out so far only for PTs of weakly coupled theories. This mechanism, however, works only for scales of the PT above about $10^8$~GeV (see e.g. also~\cite{Huang:2022vkf,Chun:2023ezg}), so that it is not testable in any laboratory experiment and only poorly so by GW telescopes, given that the predicted GW are at very high frequencies.

In this paper we point out that, if the supercooled PT is a confining one, then with respect to analogous weakly-coupled PTs, both the yield of the mass-gaining particles (i.e. the hadrons) is larger and wash-out processes are suppressed. We find that these ingredients allow to generate the baryon asymmetry down to PT scales in the TeV range, via hadron's decays. This opens up the possibility to generically test this framework with LISA and the Einstein Telescope. Depending on its specific realisation, we find that this also allows to test the scenario at colliders and/or to connect leptogenesis with neutrino mass generation in a new framework.

%This paper is organised as follows. In Sec.~\ref{sec:framework} we describe PTs with fast walls and introduce the framework; in Sec.~\ref{sec:GWs} we compute the associated GW signal. We then study in detail two specific realisations of this framework: one of baryogenesis in Sec.~\ref{sec:scalars}, where the quanta of the confining group are charged under the SM and the BAU is generated via decays of composite scalars; one of leptogenesis in Sec.~\ref{sec:leptogenesis}, where the quanta are instead SM singlets and the BAU is generated via leptogenesis from decays of composite sterile neutrinos, which also give neutrino masses via inverse see-saw. We conclude in Sec.~\ref{sec:conclusions}.

\section{Framework for the generation of the Baryon Asymmetry}
\label{sec:framework}

\subsection{Supercooled confining phase transitions}

As the temperature of the early universe bath drops, the vacuum state of the theory that is energetically favoured can change, triggering a PT. If the PT is first order it proceeds via nucleation of bubbles of the new (`true') vacuum in the bath, see e.g.~\cite{Hindmarsh:2020hop,Gouttenoire:2022gwi} for reviews.

A first-order phase transition in the early universe is controlled by the interplay between the energy density of radiation and the difference $\Delta V$ between the ones of the old (`false') and new vacuum. The latter is given by
\begin{equation}
\Delta V
\equiv \cvac\,f^4\,
\label{eq:cvac},
\end{equation}
where  $f$ is the vacuum expectation value (VEV) of the order parameter $\phi$ of the phase transition in the true vacuum (which we will later identify with the dilaton), and $\cvac$ is a dimensionless, model dependent, quantity parameterizing the vacuum energy difference, typically $\cvac\sim \mathcal{O}(1)$ or smaller. The radiation energy density is given by
\begin{equation}
\rho_\text{rad}=\frac{g_*\pi^2}{30}T^4\, ,
\end{equation}
where $g_*$ counts the effective degrees of freedom of the radiation bath.
The phase transition takes place around the nucleation temperature, $\Tnuc$, when the bubble nucleation rate per unit volume, $\Gamma$, becomes comparable to the Hubble parameter $H$, namely $\Gamma(\Tnuc)\sim H^4(\Tnuc)$. The PT is said to be supercooled if it happens when the universe is dominated by the vacuum energy rather than by radiation, $\rho_\text{rad} < \Delta V$, i.e.
\begin{equation}
\text{supercooling:} \qquad \qquad 
\Tnuc < T_\text{eq}
\equiv \Big(\frac{30\,\cvac}{g_* \pi^2}\Big)^{1/4} f\,,
\end{equation}
where $T_\text{eq}$ is defined as the temperature when $\rho_\text{rad} = \Delta V$.
%If nucleation is delayed enough, the universe, as it cools down, eventually enters a vacuum-dominated phase at the temperature
%
%\begin{equation}
%T_\text{inf}=\Big(\frac{30\,\cvac}{g_* \pi^2}\Big)^{1/4} f\, .
%\label{eq:TRH}
%\end{equation}
%
%The vacuum domination signals a period of late-time inflation, supercooling.   
Following the phase transition, the order parameter $\phi$ undergoes oscillations and decays. We assume that its width is comparable or larger than the Hubble rate, so that the universe is reheated to a temperature 
\begin{equation}
T_\RH = \Big(\frac{30\,\cvac}{g_\RH \pi^2}\Big)^{1/4} f\,,
\label{eq:TRH}
\end{equation}
%
%(g_*/g_*_\text{broken})^{\!\frac{1}{4}}\,T_\text{inf},
where $g_\RH$ is the number of degrees of freedom in the broken phase that are reheated.

A supercooled PT is generically predicted by potentials $V(\phi)$ that are flat enough at the origin at zero temperature~\cite{Witten:1980ez}, see~\cite{Levi:2022bzt} for a recent systematic study. The needed features of $V(\phi)$ are automatically predicted by models that are approximate scale invariant in the UV, and that spontaneously break scale invariance in the IR, like in radiative generation of IR scales \`a la Coleman-Weinberg~\cite{Coleman:1973jx,Gildener:1976ih}.

In this paper we will be interested in strongly coupled realisations, where confinement occurs from the condensation of an approximately conformal strongly-interacting sector, and which are dual to 5D warped models, see e.g.~\cite{Creminelli:2001th,Randall:2006py,Nardini:2007me,Konstandin:2011dr}. 
The spontaneous breaking results in a Goldstone boson parametrically lighter than $f$, the dilaton $\phi$, which can be thought of as a condensate of UV quanta of the confining theory. Its potential is generated from a small explicit breaking of scale-invariance in the UV, without which the symmetry could not be spontaneously broken, so that (see e.g.~\cite{Baldes:2021aph})

\begin{equation}
\Delta V
= V(\phi=0)-V(\phi=Zf)\simeq \Big(\frac{Z m_{\phi}f}{4}\Big)^2 \,,
\label{eq:DeltaV}
\end{equation}
where $Z$ is the dilaton's wavefunction renormalization and in strongly coupled theory is expected to be of order one. 
%This relation links the quantity $\cvac$ defined in Eq.~\eqref{eq:cvac} to the dilaton mass $m_{\chi}$ as
%
%\begin{equation}m_{\chi}= \frac{4}{Z}\,\cvac^{1/2}\,f\, .
%\end{equation}
We clarify that $\phi$ then denotes the canonically normalized dilaton, and $m_\phi$ its mass.
Eq.~(\ref{eq:DeltaV}) then implies that $m_\phi$ and $\cvac$ are not independent parameters, but are linked via
\begin{equation}
m_{\phi}
= \frac{4}{Z}\,\cvac^{1/2}\,f\,.
\label{eq:mdil_cvac}
\end{equation}
$m_\phi < $ few$\times f$ is enough to achieve values of $\Tnuc$ orders of magnitude smaller than $T_\text{eq}$, see e.g.~\cite{Baldes:2020kam,Baldes:2021aph}. Larger hierarchies between $m_\phi$ and $f$, corresponding to very small values of $\cvac$, may imply that the PT never completes. One could then envisage to complete it via the PT associated to another sector, see~\cite{Iso:2017uuu,vonHarling:2017yew} for examples where the QCD PT triggers a supercooled EW PT that would otherwise not complete. Obtaining large hierarchies between $m_\phi$ and $f$ is also non-trivial from the theoretical point of view, see e.g.~\cite{Contino:2010unpub,Appelquist:2010gy,Bellazzini:2012vz,Coradeschi:2013gda,Chacko:2013dra,Megias:2014iwa} for successful attempts.

\subsection{Dynamics of the walls}
Supercooled PTs generically predict that bubble walls are ultrarelativistic.
As bubbles are nucleated and start to expand, the wall's boost factor $\gwp$ starts growing as well because the surface tension of the wall loses against the difference in pressures. If nothing prevents their acceleration, then walls `run away' and $\gwp$ keeps growing linearly with the bubble radius~$R$. 

For the values of $\Delta V/\rho_\text{rad} \gtrsim 1$ of our interest, plasma effects cannot prevent the walls from becoming faster and faster~\cite{Konstandin:2010dm,Laurent:2022jrs}.
Particle physics effects however can. They come from particles obtaining a mass at the wall~\cite{Bodeker:2009qy} and radiation from these particles~\cite{Bodeker:2017cim}, which exert a pressure on the walls and can prevent them from running away.
In the context of PTs that are both supercooled and confining, the pressure arises from non-perturbative effects and reads~\cite{Baldes:2020kam}
%The wall boost factor in the plasma frame, $\gwp$, enters the expression of the baryon yield $Y_B$, Eq.~\eqref{eq:Yield}, therefore one should analyze the possible values that it can take during the PT.\\ 
%As bubbles are nucleated and start to expand, 
%$\gwp$ starts growing as well. If nothing slows down the bubble-wall acceleration, then $\gwp$ keeps growing as $\gwp(R)=R/(3R_{nuc})$,~\cite{Gouttenoire:2021kjv}, where $R$ is the bubble radius and $R_{nuc} \approx \Tnuc^{-1}$  is its radius at nucleation, until its value at
%the time of bubble-wall collision, $\gwp^{\text{runaway}}$. Sources of friction that could prevent this runaway regime are given by the equivalent, in this scenario, of the so-called leading order (LO) and
%next-to-leading order (NLO) contributions in~\cite{Bodeker:2009qy} and~\cite{Bodeker:2017cim}. However, in the context of supercooled confinement, the pressure $\mathcal{P}_{LO}$ arises from
%non-perturbative effects and its value was computed in~\cite{Baldes:2020kam}
%
\begin{equation}
\mathcal{P}
\simeq \frac{\zeta(3)}{\pi^2}g_\TC\,\gwp\,\Tnuc^3f\,,
\label{eq:pressure}
\end{equation}
where $g_\TC$ is the number of degrees of freedom, that are charged under the confining group, in the deconfined phase (see the next two subsections).
The non-perturbative effects responsible for this pressure can be thought of as the strong-coupling limit of the resummed radiation at the walls in weakly coupled theories, and indeed the pressure associated to the latter effects also scales linearly in $\gwp$~\cite{Gouttenoire:2021kjv}.
Walls then reach a terminal constant boost, $\gwpT$ as soon as $\Delta V = \mathcal{P}$. If this condition is never met, they run away until they collide with walls from other bubbles.
%Note that it grows linearly in $\gwp$, like the pressure from resummed radiation in weakly coupled theories~\cite{Gouttenoire:2021kjv}.
%One can observe that in confining PTs $\mathcal{P}_{LO}$ grows linearly in $\gwp$, unlike in ‘standard’ PTs where it is independent of the boost factor. The value of $\gwp$ at percolation is given by

The wall boost factor at bubbles collision then reads
\begin{equation}
\label{eq:gammawp}
\gcoll
\simeq \text{Min}\Big[\gwprun,\gwpT\Big]
\simeq \text{Min}\Big[1.7\frac{10}{\beta/H}\Big(\frac{0.01}{\cvac}\Big)^{1/2}\frac{\Tnuc}{f}\frac{M_\Pl}{f}, 10^{-3}\frac{\cvac}{0.01}\frac{80}{g_\TC}\Big(\frac{f}{\Tnuc}\Big)^3\Big]\, ,
\end{equation}
where the first entry is the runaway value at collision $\gamma_\text{run} = R_\text{coll}/(3R_\text{nuc})$~\cite{Gouttenoire:2021kjv}, $R_\text{coll}=\pi^{1/3}/\beta$~\cite{Enqvist:1991xw} is the typical bubble radius at collision and $R_\text{nuc}$, of order $\Tnuc^{-1}$, the one at nucleation.
%Since in supercooled PTs $R_\text{nuc} \lesssim \Tnuc^{-1}$ (see e.g.~\cite{Baldes:2020kam}), we use $R_\text{nuc} = \Tnuc^{-1}/2$ for definiteness.
Since in the supercooled PTs of our interest $R_\text{nuc} \lesssim \Tnuc^{-1}$~\cite{Baldes:2020kam}, we use $R_\text{nuc} = \Tnuc^{-1}/2$ for definiteness (other supercooled PTs may have $R_\text{nuc} \gtrsim \Tnuc^{-1}$, see e.g.~\cite{Baldes:2023rqv}).
We have also introduced the reduced Planck mass $M_\Pl \simeq 2.4\times 10^{18}$~GeV and $\beta \equiv (d\Gamma/dt)\Gamma$, where we remind that $\Gamma$ is the bubble nucleation rate per unit volume, so that $\beta/H$ controls the typical number of bubbles per Hubble volume and $\beta/H$ between 10 and 100 is a typical range for supercooled PTs.
%It was shown in~\cite{Baldes:2020kam} that the NLO pressure is always subleading in the parameter space of our interest, therefore $\gwp^{NLO}$ does not enter Eq.~\eqref{eq:gammawp}. Finally $\beta_H$ controls the typical number of bubbles
%per Hubble volume. $\beta_H=10$ is a typical value for supercooled PTs which we employ from now on. 
%While in the runaway regime the walls keep accelerating until bubble-walls collide, if the LO pressure equals that of the internal pressure from the difference in vacuum energies, then a terminal velocity is reached. 
Whether the bubble-walls stay in the runaway regime or reach a terminal velocity depends on the amount of supercooling of the PT, which we quantify by the ratio $\Tnuc/f$. Examining Eq.~\eqref{eq:gammawp} it is clear that, in order to have a sufficiently large $\gcoll$, one needs $\Tnuc\ll f$.

%Examining Eq.~\eqref{eq:gammawp} it is clear that in order to have a sufficiently large $\gwp$, such that the hadrons $\Psi$ can cross the wall, one needs $\Tnuc\ll f$, therefore our mechanism needs a certain amount of supercooling to produce the observed baryon asymmetry.

\subsection{Abundance of hadrons}
We call hadron any composite state of the confining sector.
Hadrons are expected to have a mass $M_\text{hadr} \approx 4\pi f \gg \Tnuc$ coming from their strong coupling to the dilaton, unless they are lighter for symmetry reasons, like the dilaton itself or other pseudo-goldstone bosons (analogous to pions in QCD).
The hadrons do not exist before the PT, where the fundamental `techniquanta' (TC) are massless (like quarks and gluons in QCD).
These are in equilibrium with the thermal bath before the phase transition and their number density to entropy ratio, the Yield, is given by
\begin{equation}
Y_\TC
\equiv \frac{n_\TC}{s}
= \frac{45 \zeta(3) g_\TC}{2\pi^4g_*}\, ,
\end{equation}
where $g_\TC$ are the degrees of freedom of the techniquanta and $n_\TC$ is their number density.

What happens upon the PT was first modeled in~\cite{Baldes:2020kam} and can be summarized as follows.
Techniquanta enter the bubble and form fluxtubes that point towards the wall. If techniquanta have more energy than the confinement scale in the wall frame, i.e. if $\gwp \Tnuc \gtrsim f$, then these fluxtubes break forming a large number of hadrons, which on average follow the bubble walls in the plasma frame\footnote{Other subtleties can change the picture as $\cvac$ and $\Tnuc/f$ approach 1, but they won't matter for the parameter space of our interest.}.
Fluxtube's breaking also results in the ejection of techniquanta outside the wall, to conserve charge. Both hadrons and ejected particles have very high energies in the bath frame, of order $\gwp f$.
As bubbles expand and collide, hadrons and ejected techniquanta collide with those from other bubbles and with particles of the pre(heated) plasma, whose typical energies are of the order of the dilaton mass $m_\phi$. These collisions result in deep inelastic scatterings which ultimately convert the large center of mass energies (of order $m_\phi \gwp f$ in the bath frame) into hadron masses.
This conversion has been found by simulations of scatterings and hadronization in~\cite{Baldes:2020kam}, and its simplicity makes it pretty robust against the details of this simulation and other possible processes affecting the propagation and scatterings of hadrons and ejected techniquanta~\cite{Baldes:2023boh}.
 This redistribution of large energies into hadron masses then enhances the yield of hadrons, with respect to that of particles that gain mass at a non-confining PT, by a factor $\gcoll f m_{\phi}/M_{\text{hadr}}^2$. The yield of a given hadron $\Psi$ then reads
%Upon the phase transition, provided that $\gwp>M_{\Psi}/\Tnuc$, such that $\Psi$ can enter the bubble, $Y_{\Psi}$ is maintained across the wall, while $n_{\Psi}^{eq}$ receives a large Boltzmann suppression, therefore the $\Psi$ particles are out-of-equilibrium inside the bubble. A baryon asymmetry can be produced by the $\Psi$ quanta inside the bubble, provided that they decay in a $B-L$ and CP violating way. \\With respect to the non-confining scenario there is enhancement by $Y^{DIS}$ of the final baryon yield, due to the presence of deep inelastic scatterings (DIS) of the techniquanta of the confining sector~\cite{Baldes:2020kam}, which increase the number of hadrons, and, via their decay into $\Psi$, effectively increase the yield. We can parametrize $Y^{DIS}$ as
%
\begin{equation}
Y_\Psi %= K^{DIS}g^{TC}\,Br\big(\text{had}\to \Delta\big)\,
\simeq Y_\TC
\times 3\frac{\gcoll f m_{\phi}}{M_{\text{hadr}}^2}\, \,Br\big(\text{hadr}\to \Psi\big)
\times\, \Big(\frac{\Tnuc}{T_\text{eq}}\Big)^{\!3} \,\frac{T_\RH}{T_\text{eq}},
\label{eq:Ypsi}
\end{equation}
%
%where $K^{DIS}$ is the average number of hadrons produced via the scatterings by a single resonance, we take $m_{\text{hadr}}\sim 4\pi f$, $g^{TC}\sim 100$ denotes the degrees of freedom of the quanta of the confining sector and $Br\big(\text{had}\to \Psi\big)$ is the branching ratio of the hadrons of the confining sector into the $\Psi$. 
where $M_{\text{hadr}} \simeq 4 \pi f$ and $Br\big(\text{hadr}\to \Psi\big) < 1$ is the number of hadrons $\Psi$ produced over all hadrons in the end of the scattering chains, which is expected to be suppressed if $M_\Psi \gg f$, analogously to QCD. The last factor $(\Tnuc/T_\text{eq})^3 \,T_\RH/T_\text{eq} \propto (\Tnuc/f)^3$ is the usual dilution of relics from supercooled PTs, the factor $(3\gcoll f m_{\phi}/M_{\text{hadr}}^2)$ is its enhancement specific of confining theories discussed before Eq.~(\ref{eq:Ypsi}).

\subsection{Baryon asymmetry from hadrons decay}
%We thus obtain the following expression for the baryonic yield:
%
%\begin{equation}
%\frac{Y_B}{Y_B^{\text{obs}}}
%=\epsilon_{\Psi}\,K_{Sph}\,\frac{Y_\Psi}{Y_B^{\text{obs}}}\,\Big(\frac{\Tnuc}{T_{RH}}\Big)^{3}\,Y^{DIS} \, ,
%\label{eq:Yield}
%\end{equation} 
%
%where $\epsilon_{\Psi}$ is the average baryon number produced in $\Psi$ or $\Psi^*$ decays, $K_{Sph}=28/79$ is the sphaleron reprocessing
%factor, which has to be included when the electroweak sphalerons are in thermal equilibrium, namely when $T \geq  T_\text{Sph}=132$ GeV and the $(\Tnuc/T_{RH})^3$ suppression comes from the entropy production from reheating following the phase transition.
%We point out that the size that the combination $g^{TC}Br\big(\text{had}\to \Psi\big) $ can take is model dependent, still, because of the enhanced production of hadrons via the DIS and the suppression of washout effects, the connection between $\Psi$ and a confining sector extends significantly the allowed parameter space for the generation of the baryon asymmetry with respect to the deconfined case.
%Moreover, one generally expects all the hadron masses to be of $\mathcal{O} (f)$, therefore it is natural to assume that the rate asymmetry $\epsilon_{\Psi}$ can receive an enhancement  due to some degeneracy among the hadron masses.

Let us now assume that a portal connects the confining sector and the SM that results in an otherwise-stable hadron $\Psi$ to decay violating $B$, $C$ and $CP$. Note that the portal needs not to necessarily violate $C$ and $CP$, as their violation could rely on the one existing in the SM (see e.g.~\cite{Bruggisser:2018mus,Bruggisser:2022rdm} for a realisation of this idea in composite Higgs models, although for slow bubble walls) or it could be a property of the strong sector. 
We do not need to specify this nor the UV origin of the portal in this paper, we just need to know that each of the $\Psi$ or $\bar{\Psi}$ produces on average a net baryon number $\epsilon_{\Psi}$. The baryon yield then reads
\begin{align}
Y_B
= &~\epsilon_{\Psi}\,K_\text{Sph}\,Y_\Psi \\
\simeq &~Y_\BAU \times \frac{\gcoll}{\gwpT} \,\frac{\epsilon_{\Psi}\,K_\text{Sph}}{1.2\cdot10^{-4}}
 \frac{Br\big(\text{hadr}\to \Psi\big)}{10^{-3}\, Z} \Big(\frac{107.75}{g_\RH}\Big)^{\!\frac{1}{4}} \Big(\frac{\cvac}{0.01}\Big)^{\!\frac{3}{4}} \Big(\frac{4 \pi f}{M_\text{hadr}}\Big)^2,
\label{eq:Yield}
\end{align}
where $K_\text{Sph}$ is the sphaleron reprocessing factor, $K_\text{Sph}=1$ if the baryon asymmetry is generated when electroweak sphalerons are not active, namely when $T \lesssim  T_\text{Sph}=132$ GeV~\cite{Kuzmin:1985mm}, and $K_\text{Sph}<1$ otherwise ($K_\text{Sph} = 28/79$ in the SM~\cite{Khlebnikov:1988sr}).
In general $\epsilon_{\Psi}$ is at most of order 
\begin{equation}
\epsilon_{\Psi} \lesssim \frac{\text{Im}y^2}{4\pi} \frac{M_\Psi}{\Delta M}\,,
\label{eq:eps_general}
\end{equation}
where by Im$y^2$ we mean the imaginary part of some combination of couplings with coupling dimension 2, and where we include possible enhancements from small hadron mass splittings $\Delta M \ll M_\Psi$. Small hadron mass splittings are actually expected from confining sectors with at least two light quarks (think for example of the neutron and proton masses in QCD).

The discussion so far assumes that the population of hadrons $\Psi$ is unaffected before they decay. First, we estimate the total decay rate of $\Psi$ as $\Gamma_{\Psi} = C|y|^2 M_{\Delta}/8\pi$, where $C$ is a model-dependent factor, and observe that they are much faster than the Hubble rate in the entire parameter space that we will consider. So we should compare the rate of $\Psi$ annihilations, i.e. how fast the $\Psi$ want to go to their thermal equilibrium abundance, with $\Gamma_{\Psi}$  and not with $H$.
$\Psi$'s annihilate into dilatons $\phi$ and possibly other particles, depending on the model. As the coupling of $\Psi$ to dilatons is model-independent and large (of order $M_\Psi/(Zf)$), we estimate the annihilation cross section with the analogous QCD one for $p-\bar{p}$ annihilations, $\sigma(\Psi \Psi^{*} \to \phi \phi) v_{rel} \sim 4\pi/M_\Psi^2$.
Therefore, using Eq.~\eqref{eq:Yield}, we obtain the total $\Psi$ annihilation rate
\begin{equation}
\label{eq:annhPsi}
\Gamma_{\text{annh}}
\sim n_\Psi \sigma v_{rel}
\sim 8.7 \times 10^{-8} M_{\Psi} \Big(\frac{4\pi f}{M_\Psi}\Big)^{\!5} \frac{\gcoll}{\gwpT} 
 \frac{Br\big(\text{hadr}\to \Psi\big)}{0.1\, Z} \Big(\frac{\cvac}{0.01}\Big)^{\!\frac{3}{2}}\frac{107.75}{g_\RH}\frac{g_*}{200}\, .
\end{equation}
In order to not change the abundance of the $\Psi$s before they decay and source the BAU, we require $\Gamma_{\text{annh}}$ to be below $\Gamma_{\Psi} = C {|y|^2} M_{\Delta}/8\pi$, i.e. the following ratio should be smaller than one:
\begin{equation}
\label{eq:annhoverdecayPsi}
\frac{\Gamma_{\text{annh}}}{\Gamma_{\Psi}}
\simeq \frac{0.02}{C} \times \Big(\frac{0.01}{|y|}\Big)^2\Big(\frac{4\pi f}{M_\Delta}\Big)^{\!5} \frac{\gcoll}{\gwpT} 
 \frac{Br\big(\text{hadr}\to \Psi\big)}{0.1\, Z} \Big(\frac{\cvac}{0.01}\Big)^{\!\frac{3}{2}}\frac{107.75}{g_\RH}\frac{g_*}{200} \, .
\end{equation}
In the regions where this is larger than one, it could still possible to source the BAU, as there is still a population of $\Psi$s.
The determination of the BAU would require additional calculations.
Since in most of the parameter space of our interest $\Gamma_{\text{annh}}<\Gamma_{\Psi}$, for simplicity we do not compute the BAU there but just warn the reader whenever we have $\Gamma_{\text{annh}}>\Gamma_{\Psi}$.

 \begin{figure}
\centering
  \begin{minipage}[b]{0.48\textwidth}
   \includegraphics[height= 7cm, width=8cm]{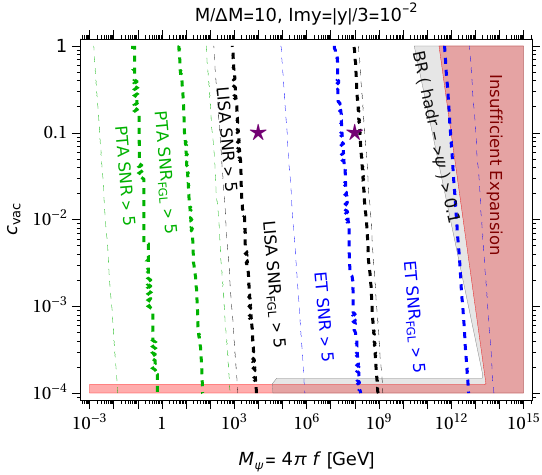}
    \end{minipage}
    \vspace{0.5cm}
   \hfill
 \begin{minipage}[b]{0.48\textwidth}
\includegraphics[height= 7cm, width=8cm]{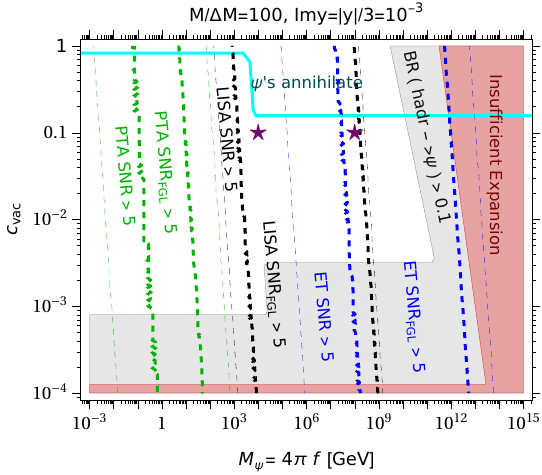}
  \end{minipage}  
  \caption{
  Parameter space where our baryogenesis framework manages to reproduce the observed BAU, and its testability in gravity waves with LISA, the Einstein Telescope and the PTA.
We choose $g_\TC = 80$, $g_* = g_\SM + g_\TC$ and, for definiteness, $g_\RH = g_\SM + 1$ (where the +1 accounts for the dilaton), the dilaton wavefunction normalization $Z=1$ and $\Tnuc/f=10^{-3}$.
The white area is allowed by the PT dynamics and can accomodate the baryon asymmetry, where we saturate the baryon yield at each point,
 by obtaining $ Br \big(\text{hadr}\to \Psi \big)$ from Eq.~(\ref{eq:Yield}). 
The excluded gray region presents values of $ Br \big(\text{hadr}\to \Psi \big)>0.1$, chosen as a benchmark upper limit, hence it cannot account for the whole baryon asymmetry of the Universe. The excluded red region presents $\gcoll<M_{\Psi}/\Tnuc$, hence the hadrons cannot enter the bubble and provide the BAU. Above the cyan lines hadrons start to equilibrate with dilatons before they decay, so the computation of the BAU should be different than the one we employed. 
The regions delimited by the green, black and blue dashed lines are testable, respectively, by PTAs, LISA and the Einstein Telescope, where the astrophysical foregrounds have (not) been taken into account in the thicker (thinner) lines. We denote with violet stars the two points of the parameter space for which we computed the gravitational wave spectrum shown in Fig.~\ref{fig:gwspectrum}.
The viable parameter space will be reduced in specific realizations of our framework, by model-dependent washout effects or laboratory searches depending on the SM interactions of the hadrons $\Psi$, see Figures~\ref{fig:parspace} and~\ref{fig:parspaceISS}.
}\label{fig:parspace_framework}
\end{figure}

The parameter space where our framework can potentially reproduce the observed BAU is displayed in Fig.~\ref{fig:parspace_framework} for representative benchmark values of the parameters, for $f \geq$~MeV cause lower values are excluded by Big Bang Nucleosynthesis (BBN), see e.g.~\cite{Gouttenoire:2023bqy}.
Imposing $Y_B = Y_\BAU$ fixes the value of $Br\big(\text{hadr}\to \Psi\big)$ in every point of the parameter space. We require $Br\big(\text{hadr}\to \Psi\big) < 0.1$ because string fragmentation is expected to give rise mostly to light hadrons, and at least one light hadron, the dilaton, exists in our picture. We also require $\Gamma_{\text{annh}}/\Gamma_{\Psi} < 1$, for definiteness for $C=1$ and using $|y|$=3 Im$y$.
%, and equating $\epsilon_\Psi$ to its upper limit of Eq.~(\ref{eq:eps_general}). 
As long as the walls are in the terminal velocity regime, $Y_B$ is independent of both $f$ and $\Tnuc/f$. These parameters play a role as soon as collisions happen in the runaway regime, i.e. in a region of large $f$ and small $\Tnuc/f$, where $Y_B$ gets quickly suppressed (indeed Fig.~\ref{fig:parspace_framework} displays that one cannor reproduce the BAU at large $f$). 

$f$ and $\Tnuc/f$ also play a role in defining the regions where the BAU %can be washed out by various effects, which are not included in Eq.~(\ref{eq:Yield}) because they depend on the model realising our framework. One can still identify general expectations for washout processes.
can be washed out by various effects. Washout effects are not included in Eq.~(\ref{eq:Yield}), nor they are visualized in Fig.~\ref{fig:parspace_framework}, because they depend on the specific model realising our framework. One can still identify the following general expectations for them.
Concerning inverse decays, they can be neglected  as long as $M_\Psi \gg T_\RH$, because their rate is Boltzmann-suppressed at large $M_\Psi/T_\RH$. $M_\Psi \gg T_\RH$ is actually a consequence of $T_\RH \lesssim f$ (see Eq.(\ref{eq:TRH})) and $M_\Psi \simeq 4\pi f$.
$M_\Psi/f$ is the hadron coupling to the dilaton, the large values needed to avoid washout from inverse decays can only be obtained in strongly-coupled theories.
Other washout processes that generically exist are $2\to2$ scatterings mediated by $\Psi$.
Requiring that they are negligible imposes i) an upper bound on Im$y^2$ of Eq.~(\ref{eq:eps_general}), hence our normalization of $\epsilon_{\Psi}$ in Eq.~(\ref{eq:Yield}) to a small number, and ii) an absolute lower limit on $M_\psi$, because they are suppressed by some power of $M_\psi$. Condition ii) is what prevents from going to values of $f$ much below a TeV, although it is possible that a minimal model-building effort weakens this requirement.
Another obstruction to achieving $f \lesssim$~TeV arises in models where the confining sector generates a lepton asymmetry, because then one needs $T_\RH > T_\text{Sph}$ to transmit it to the baryon sector.

Two key features of our framework, that are different with respect to its weakly coupled analogue of~\cite{Baldes:2021vyz,Chun:2023ezg}, can already be grasped by our general discussion so far:
\begin{itemize}
\item[$\diamond$] the hadron yield is enhanced by the strong dynamics and it allows for a larger BAU;
\item[$\diamond$] the strong coupling $M_\Psi/f \gg 1$ suppresses washout coming from inverse decays.
\end{itemize}
They imply that this mechanism works for scales of the PT down to (at least) the weak scale, which opens the possibility to test it with both LISA and the Einstein Telescope, which is the topic of Sec.~\ref{sec:GWs}.
It also makes the mechanism testable in the laboratory, either directly producing the hadrons at high-energy colliders, or indirectly testing their $CP$ and flavour violating effects.
Finally, a scale of the PT down to (at least) the TeV scale also allows for interesting theoretical connections, e.g. with models explaining the origin of the weak scale, or of neutrino masses, with confining dynamics. We will explore some of these consequences in two explicit models that realise our framework, in Secs.~\ref{sec:scalars} and~\ref{sec:leptogenesis}.

\section{Gravitational wave signal}
\label{sec:GWs}

We now detail the expected GW signal from a supercooled PT.
Using the numerical results from~\cite{Cutting:2020nla}, the GW energy density power spectrum as measured today, for initially thick walled, runaway bubbles is given by 
\begin{equation}
\label{eq:runspectrum}
h^2 \Omega_\GW (\nu)
\equiv h^2 \frac {d\,\Omega_\GW}{d\,\text{log} (\nu)}
= h^2 \Omega_\text{rad}^0 \,  \Big(\frac{g_0}{g_\RH}\Big)^{\!\frac{1}{3}} \,\Omega_\GW^\PT (\nu)
\simeq 4.4 \times 10^{-7} \Big(\frac{100}{g_\RH}\Big)^{\!\frac{1}{3}}
\Big(\frac{\alpha_\GW}{1+\alpha_\GW}\Big)^{\!2} \frac{S_\phi (\nu)}{(\beta/H)^2}\,,
\end{equation}
where $h\simeq 0.67$ is the dimensionless Hubble parameter as determined by Planck~\cite{Planck:2018vyg}.
In Eq.~(\ref{eq:runspectrum}) the factors $\Omega_\text{rad}^0$ and $(g_0/g_\RH)^{\frac{1}{3}}$ take into account that the GW energy density redshifts as radiation from its value at the PT, $\Omega_\GW^\PT$, until today, and $\Omega_\text{rad}^0$ and $g_0$ are, respectively, the energy fraction in radiation and the degrees of freedom in entropy today.
$\alpha_\GW$ is the energy released as bulk motion during the PT (which we approximate as the energy difference between the false and true vacuum, $\Delta V$) normalized to the radiation density $\rho_\text{rad}$, namely
\begin{equation}
\alpha_\GW
\simeq \frac{\Delta V}{\rho_\text{rad}}
= \Big(\frac{T_\text{eq}}{\Tnuc}\Big)^4
= \frac{30\cvac}{\pi^2 g^{*}} \Big(\frac{f}{\Tnuc}\Big)^{4},
\end{equation}
thus one expects $\alpha_\GW\gg 1$ for supercooled PTs. The shape of the spectrum is governed by the spectral function 
\begin{equation}
S_\phi (\nu)=\frac{(a+b)\tilde{\nu}^b \nu ^a}{b\,\tilde{\nu}^{(a+b)}\,+a \,\nu ^{(a+b)}} \,,
\end{equation}
where we adopt the central values for the bubbles in the setup with the thickest walls from~\cite{Cutting:2020nla}, namely $\{a,b\}=\{0.742,2.16\}$. The peak frequency of the signal today is given by 
\begin{equation}
\label{eq:runpeak}
\tilde{\nu}= 33 \,\mu\text{Hz}\,\Big(\frac{g_\RH}{100}\Big)^{1/6} (\beta/H) \Big(\frac{T_\RH}{10^3 \text{ GeV}}\Big)\,.
\end{equation}
Finally, one should impose that $\Omega_\GW \propto \nu^3$ at small frequencies, due to causality for super-horizon modes at GW production \cite{Durrer:2003ja}. We therefore apply a cut in the spectrum at the redshifted frequency today of
\begin{equation}
\label{eq:cut}
\nu^{*}(t_0)=\frac{a (T_\RH)}{a_0}\, \frac{H(T_\RH)}{2\pi} \, ,
\end{equation}
below which we enforce the correct IR scaling and where $a(T_\RH)/a_0$ is the ratio of the scale factors between the PT and today.

The numerical simulations performed in~\cite{Cutting:2020nla} assume that bubbles effectively runaway until collision, however a large part of our parameter space, notably the one associated to a PT around the TeV scale, sits in the terminal velocity regime, where the vacuum energy pressure is counterbalanced by the friction produced by particles in the plasma. The vacuum energy is thus partly converted into kinetic energy of the particles in the plasma already prior to the bubble wall collision. However, the contribution from sound waves or turbulence~\cite{ Caprini:2015zlo,Caprini:2019egz} in supercooled PTs is not yet clearly understood. Indeed, current hydrodynamical simulations, which aim to capture the contribution of the bulk motion of 
the plasma to the GW signal, do not yet extend into the regime in which the energy density in radiation is subdominant to the vacuum~\cite{Cutting:2019zws} and analytical studies of shock-waves in
the relativistic limit have just started~\cite{Jinno:2019jhi}. In this scenario one can argue
that derivatives of the envelope approximation such as the so-called bulk flow model are more realistic in modelling the GW signal.

\begin{figure}[!hbt]
  \centering
  \includegraphics[width=0.6\linewidth]{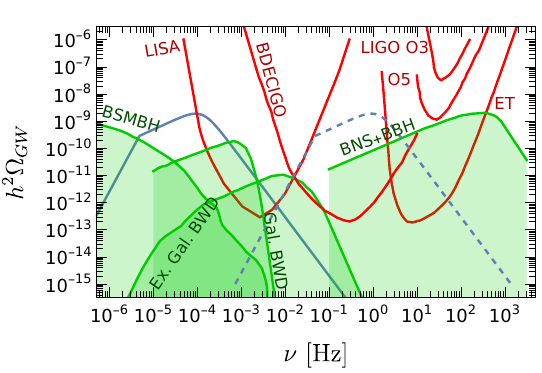} 
  \caption{
  Example of GW spectra for two benchmark points of the $\{\cvac,M_{\Psi}\}$ parameter space. The solid (dashed) blue line shows the predicted gravitational wave spectrum for the PT, with $\{\cvac,M_{\Psi}\}=\{10^{-1},10^4 \text{ GeV}\}$ and $\{\cvac,M_{\Psi}\}=\{10^{-1},10^8 \text{ GeV}\}$. We have assumed $\Tnuc/f=10^{-3}$, $\alpha_\GW=100$ and $\beta/H=15$ as typical benchmark values for supercooled PTs. The spectra are compared with PLI sensitivity curves, shown as red lines, with signal-to-noise ratio SNR=10 for LISA~\cite{Caprini:2019pxz} and the Einstein Telescope (ET)~\cite{Branchesi:2023mws}.
  We also show PLI sensitivity curves for LIGO-VIRGO O3, LIGO-VIRGO Design A+ sensitivity~\cite{KAGRA:2021kbb} and BDECIGO~\cite{Kawamura:2020pcg}. Estimated astrophysical foregrounds from binary super-massive black holes~\cite{Rosado:2011kv}, galactic white-dwarf binaries~\cite{Robson:2018ifk}, extragalactic white-dwarf binaries~\cite{Farmer:2003pa}, binary black holes and neutron stars~\cite{KAGRA:2021kbb} are also shown as green shaded regions.}
  \label{fig:gwspectrum}
\end{figure}

In practice, the difference between~\cite{Cutting:2020nla} and bulk-flow models like~\cite{Konstandin:2017sat} is very mild, see e.g. the comparison in~\cite{Baldes:2023rqv}.
Therefore we use the parametrization from~\cite{Cutting:2020nla}, described above, for both regimes of wall velocities.
Examples of GW spectra are shown in Fig.~\ref{fig:gwspectrum}, for $\alpha_\GW = 100$ and $\beta/H=15$, both typical of supercooled PTs and for two benchmark values of $\cvac$ and $M_\Psi$.

Our model can be probed by several current or upcoming GW interferometers, due to the wide parameter space allowed.
%Therefore we focus on signals that can be tested at the Einstein Telescope (ET) and LISA, as they are sensitive to the range of frequencies typically associated with the scale of the PT in our setup.
The sensitivity of a detector to a GW signal is given by 
\begin{equation}
\Omega_{\text{sens}} (\nu)=\frac{4\pi^2}{3 H_0^2} \nu^3 S_n (\nu)\, ,
\end{equation}
where $H_0 = 100 h$ km/s/Mpc is the present day Hubble parameter and $S_n$ is the noise spectral density~\cite{Hild:2010id}.
The signal to noise ratio SNR to a GW background is given in terms of the above quantities as~\cite{Thrane:2013oya} 
\begin{equation}
\label{eq:snr}
\text{SNR}=\sqrt{t \int_{\nu_{\text{min}}}^{\nu_{\text{max}}} d\nu \Big(\frac{\Omega_\GW (\nu)}{\Omega_{\text{sens}} (\nu)}\Big)^2} \, ,
\end{equation}
where $\nu_{\text{min}}$, $\nu_{\text{max}}$ and $t$ denotes the minimal and maximal frequencies accessible at the detector and the fiducial observation time, which is taken to be $t$= 10 years for the ET and $t$=3 years for LISA.
Using this expression, the so called power-law-integrated (PLI) sensitivity curves are computed~\cite{Thrane:2013oya}, 
%and our calculations for the ET~\cite{Hild:2010id} and LISA~\cite{Caprini:2019pxz} PLI sensitivity curves are shown in Fig.~\ref{fig:gwspectrum}, for a fiducial choice of SNR=10.
and our calculations for the PLI sensitivity curves of, LISA~\cite{Caprini:2019pxz}, BDECIGO~\cite{Kawamura:2020pcg}, LIGO-VIRGO-KAGRA~\cite{KAGRA:2021kbb} and Einstein Telescope (ET)~\cite{Branchesi:2023mws} (configuration `2L $0^\circ$ 20km' for definiteness) are shown in Fig.~\ref{fig:gwspectrum}, for a fiducial choice of SNR=10.

There are several astrophysical stochastic GW foregrounds
% in the ET and LISA frequency range
which could mimic the GW signal coming from the PT, see Fig.~\ref{fig:gwspectrum} for details. In order to take this limitation into account we also define a foreground-limited signal to noise ratio $\text{SNR}_{\text{FGL}}$,
\begin{equation}
\label{eq:snrastro}
\text{SNR}_{\text{FGL}}=\sqrt{T \int_{\nu_{\text{min}}}^{\nu_{\text{max}}} d\nu \Big(\frac{\text{Max}[\Omega_{GW} (\nu)-\Omega_{FG}(\nu),0]}{\Omega_{\text{sens}} (\nu)}\Big)^2} \,.
\end{equation}
With the above conservative definition we impose that our signal is not detectable if buried under the astrophysical foreground $\Omega_{FG}(\nu)$.
We show in Fig.~\ref{fig:parspace_framework}
%s.~\ref{fig:parspace} and~\ref{fig:parspaceISS}
the regions of parameter space of our BAU framework that could be probed by Pulsar Timing Arrays (PTA), LISA and the ET, which we define as those where $\text{SNR}_{\text{FGL}} \geq 5$. 
For comparison, we also show the regions that one could probe in absence of astrophysical foregrounds, which could be relevant in case one will envisage ways to tell them apart from a GW signal from PTs. The two sensitivities are anyway very similar for LISA, due to the strong GW signal from the PT in comparison with foregrounds,  while for PTAs and ET we find an appreciable difference.
The sensitivity showing up for $f \lesssim 100$~GeV is the one attainable using GWs as measured by PTA, which we compute following~\cite{Aggarwal:2018mgp} and identifies yet another way to test our framework, in case one manages to make it work down to such small scales of the PT.
If one did, then one could potentially explain the GW background observed by PTAs~\cite{NANOGrav:2020bcs,Goncharov:2021oub,EPTA:2021crs,Antoniadis:2022pcn} with a PT~\cite{NANOGrav:2021flc,Bringmann:2023opz} that, at the same time, explains the BAU via the framework proposed in this paper.
%One can indeed observe that a very large fraction of the allowed space could indeed be probed by the GW interferometers.
In any case, a very large fraction of the allowed space could indeed be probed by GW interferometers.
%An example of GW spectrum is shown in Fig.~\ref{fig:gwspectrum} for two benchmark points in the $\{\cvac,M_{\Psi}\}$ space, where we show explicitly that, for typical values of the parameters $\alpha_\GW$ and $\beta/H$ in case of supercooled PTs, one expects a GW signal to be in the reach of the planned experiments. 

\section{Model 1: Baryogenesis from a composite scalar}
\label{sec:scalars}
We have so far discussed a general framework for the generation of the baryon asymmetry and we have established that it presents some interesting features, as an enhanced production of the baryon yield with respect to the non-confining case and a gravitational wave signal potentially in reach of current and future GW interferometers.
%Therefore we want to implement our paradigm in a couple of specific models already discussed in the literature and show explicitly that we can extend significantly their allowed parameter space for a successful baryogenesis/leptogenesis.
We now specify our framework to two models already discussed in the literature, compute their predictions for washout and for the BAU as well as for detection, and show explicitly that we can extend significantly their allowed parameter space for a successful baryogenesis/leptogenesis.

\subsection{Model and baryon asymmetry}
We start by revisiting the model of baryogenesis discussed in~\cite{Baldes:2021vyz}, where the baryon number is produced by the $CP$-violating decays of a heavy scalar denoted there as $\Delta$. In our setup $\Delta$ is a composite scalar field originating from a dark confining sector and it is a realization of the $\Psi$ hadron discussed above. We introduce $i=1,2$ generations of the $\Delta_i$ scalars, with $M_{\Delta_2}>M_{\Delta_1}$.
 We assume these fields to transform as $(3\,,1\,,2/3)$ under the SM gauge group $\mathrm{SU(3)_c} \times \mathrm{SU(2)_L} \times \mathrm{U(1)_Y}$. We also introduce a SM gauge singlet fermion $\chi$ with mass $M_\chi$ and the following baryon number violating Yukawa interactions of $\Delta_i$:%, which are allowed by the gauge symmetry: 
\begin{equation}
\label{eq:lagrangiantriplet}
\mathcal{L} \supset y_{di}\Delta_i\,\overline{d_R^c}\,d_R\,+\, y_{ui}\Delta_i\,\overline{\chi_R}u_R^c+ \hc \,,
\end{equation}
where colour %and flavour
indices have been suppressed for simplicity of the notation.
%We observe that $y_{di}$ is antisymmetric in flavour
$y_{di}$ is antisymmetric in flavour $d$
as a consequence of the antisymmetry of the colour indices. A necessary condition for $CP$ violation is that the above couplings should be complex. Some, but not all the phases, even in the minimal scenario, can be removed through field rephasings $f\to e^{i\alpha_f}f$. To be more concrete let's examine some specific scenarios: 
\begin{itemize}
\item[$\diamond$]
In the minimal scenario there is only one copy of $\chi$ and the $\Delta_{1,2}$ couple each to one up-type quark and to a pair of down-type quarks (that do not have the same flavour). This results in 4 Yukawa couplings $y$. One can remove 3 out of their 4 phases by rephasing the $\Delta_i$ and $\chi$, but one phase survives since in general $\text{Arg}(y_{d2})-\text{Arg}(y_{d1})\neq \text{Arg}(y_{u2})-\text{Arg}(y_{u1})$. Thus in this setup there are 4 couplings and 1 (physical) phase.   
\item[$\diamond$]
One can switch on all the allowed SM quark Yukawa couplings, keeping a single copy of $\chi$. This setup features 12 couplings, 6 involving down-type quarks and 6 involving the up-type quarks. As in the previous case we can rephase the 3 new fields, so we're left with 9 independent new phases.
%We have already used all the freedom of the new fields to absorb the 3 phases as described above, however one can absorb 4 more phases by performing $U(1)$ transformations on the quark fields, with different phases for each flavour in general. Therefore in this setup there are 12 couplings and 5 phases. However it has to be pointed out that rephasing the quarks doesn't leave the charged current of the SM invariant, hence the CKM matrix is not in its canonical form but will have 5 physical phases, instead of the 1 in the canonical form. Therefore the net effect of quark rephasing is to trade the CP violation in the Yukawa couplings of $\Delta$ for CP violation in the CKM.
%
\item[$\diamond$]
For 3 copies of $\chi$, we find 24 couplings (6 from down- and 18 from up-type quarks). With respect to the previous setup we have the freedom to remove 2 more phases by redefining $\chi_{2,3}$, therefore we end up with 19 independent phases.
\end{itemize}
Note that in the second and third case (but not the first one) one could remove further phases from the Yukawa couplings of Eq.~\eqref{eq:lagrangiantriplet} by rephasing the right-handed quarks, but those phases would not be unphysical as they would show up in the quark Yukawas.

 \begin{figure}[!hbt]
  \centering
  \includegraphics[width=0.6\linewidth]{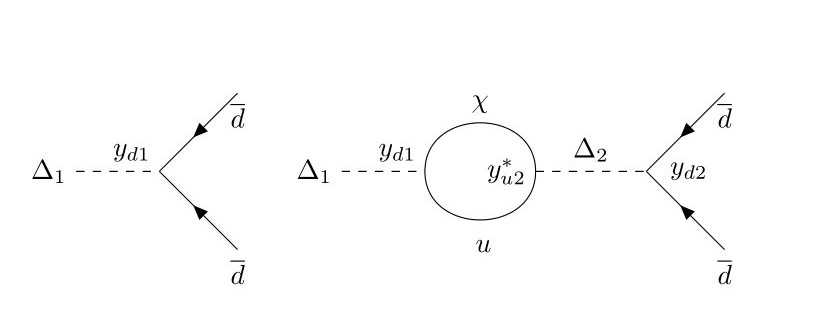}
  \caption{Example of the diagrams which interfere and lead to CP
violation. The intermediate loop particles are kinematically able to
go on-shell.}\label{fig:interference}
\end{figure}

Assuming $M_{\chi} \ll M_{\Delta_i}$ to avoid complications with phase space effects, the couplings of the Lagrangian Eq.~\eqref{eq:lagrangiantriplet}  lead to the following tree level decay rates for the scalars $\Delta_i$
\begin{equation}
\Gamma (\Delta_i \to \overline{d_R d'_R} \,)\simeq \frac{|y_{di}|^2}{8\pi}M_{\Delta_i} ,
\end{equation}
\begin{equation}
\Gamma (\Delta_i \to \chi u_R   )\simeq \frac{|y_{ui}|^2}{16\pi}M_{\Delta_i} ,
\end{equation}
where
by $d'_R$ we denote a right down quark of different flavour than $d_R$, we have not summed over flavours and we have summed over final state colours for the first decay.
%where  we have summed over the final state colours for the first decay, but we have left the summation over flavours implicit.\\
Interference between tree- and loop-level diagrams, as those shown in  Fig.~\ref{fig:interference}, leads to $CP$ violation in the decays. Focusing on the decays of  $\Delta_1$ only, $CP$ violation can be parametrized as
\begin{equation}
\begin{split}
\Gamma (\Delta_1 \to \overline{d_Rd'_R})    & =\Gamma_{1d} (1+\epsilon_d) \,,\\
\Gamma (\Delta^*_1 \to d_Rd'_R)  		   &=\Gamma_{1d} (1-\epsilon_d) \,,\\
\Gamma (\Delta_1 \to \chi u_R )   		   &=\Gamma_{1u} (1+\epsilon_u)\,, \\
\Gamma (\Delta_1^* \to \overline{\chi u_R}) &=\Gamma_{1u} (1-\epsilon_u) \,.
\end{split}
\end{equation}
Since the total decay widths of $\Delta_1$ and $\Delta_1^*$ have to be equal due to the $CPT$ theorem, it follows that $\epsilon_d \Gamma_{1d}=-\epsilon_u \Gamma_{1u}$.
The imaginary part of the loop in  Fig.~\ref{fig:interference} controls the rate asymmetry between $\Delta_1$ and $\Delta_1^*$ decays and can be extracted via Cutkosky rules \cite{1960JMP.....1..429C}. One then finds for the rate asymmetry
\begin{equation}  \label{eq:asymmetry}
 \epsilon_{\Delta}=\frac{1}{2\pi} \frac{\IM (y_{d1}^{*}y_{u1}y_{u2}^{*}y_{d2})}{|y_{ui}|^2+2|y_{di}|^2}\frac{M^2_{\Delta_1}}{M^2_{\Delta_2}-M^2_{\Delta_1}} \sim \frac{\IM [y^2]}{6\pi} \frac{M^2_{\Delta_1}}{M^2_{\Delta_2}-M^2_{\Delta_1}} \, ,
\end{equation}
where from now on we assume that there is no major hierarchy among the Yukawa couplings $y$ and we take generic $\mathcal{O}(1)$ phases. We are also discarding resonant enhancements of the rate asymmetry, namely we are assuming that $M_{\Delta_2}-M_{\Delta_1} \gg \Gamma_{\Delta_i}$. 
Finally, the baryon asymmetry is obtained by plugging $\epsilon_\Delta$ of Eq.~(\ref{eq:asymmetry}) in $Y_B$ of Eq.~(\ref{eq:Yield}).

\subsection{Washout processes}
In our scenario the most constraining washout processes are those occurring after reheating, after the PT.
The reason is that the population of the $\Delta$ hadrons, yielding the baryon asymmetry via their decays, is mostly sourced by the deep inelastic scatterings of the hadrons (produced upon string fragmentation of incoming or ejected techniquanta) with other hadrons or bath particles. 
These scatterings mostly happen after bubble percolation, so that washout effects just after wall crossing, like those considered in~\cite{Baldes:2021vyz}, are not relevant in our framework (but see the end of the previous subsection for $\Delta$ annihilations after percolation).

After bubbles percolate and all hadrons are produced, diquark interactions violating $B-L$ can result in washout if $T_\RH$ is too high. Off-shell $2\to2$ scattering processes mediated by $\Delta$ that violate baryon number, e.g. $d\,d'\leftrightarrow \overline{\chi\,u}$, have a rate that can be estimated as
\begin{equation}
\Gamma_{WO}\simeq \frac { y^4\,T^5_\RH}{8\pi M_\Delta^4} \, ,
\end{equation}
where we are integrating out $\Delta$ because $T_\RH \ll M_{\Delta}$ in our framework.
This rate is harmless for our scenario provided that it is below the Hubble rate $H$, namely
\begin{equation}
\frac{M_\Delta}{T_\RH}\geq \Big(\frac{\sqrt{90}y^4 M_\Pl}{8\pi^2 g^{*^{1/2}}T_\RH}\Big)^{1/4}\,.
\label{eq:offshell}
\end{equation}
Moreover, also inverse decays into on-shell $\Delta$ can lead to $B-L$ washout. The Boltzmann suppressed rate is given by
\begin{equation}
\Gamma_{ID}
\simeq \frac{3\,y^2}{16\pi}M_{\Delta}\Big(\frac{M_{\Delta}}{T_\RH}\Big)^{3/2}\text{Exp}\Big[-\frac{M_{\Delta}}{T_\RH}\Big]\,.
\end{equation}
This is safely below $H$, provided that
\begin{equation}
\frac{M_\Delta}{T_\RH}\geq \text{Log} \Big[\frac{y^2 M_\Pl}{8\pi\,g^{*^{1/2}}\,T_\RH}\Big(\frac{M_{\Delta}}{T_\RH}\Big)^{5/2}\Big]\,.
\label{eq:onshell} 
\end{equation}

\subsection{Proton decays and further washout from $\chi$}
So far we have not addressed the detailed nature of $\chi$, which we assume not to be linked to the confining sector, to allow us more freedom with its phenomenology. A first idea is to take $\chi$ to be close to massless, which would be the case if the SM neutrinos gain  Dirac mass via a Yukawa-type interaction $y_{\nu}\, \overline{L} H \chi_R$, with $y_{\nu}\sim 10^{-12}$.  However, $\chi$ allows nucleon decays via t-channel diagrams mediated by $\Delta$, like $p \to K^{+} \chi$, as long as $M_\chi<m_p -m_K^{+}$. Taking the couplings to all the generations to be of the same order, we estimate the decay rate
\begin{equation}
\Gamma (p\to K^{+} \chi)
\sim \frac {y_u^2 y_d^2}{64 \pi^3} \frac{m_p^5}{M_\Delta ^4}
\simeq \Big(\frac{y}{10^{-2}}\Big)^4 \Big(\frac{\text{TeV}}{M_{\Delta}}\Big)^{4}10^{-23}\text{ GeV}.
\label{eq:proton}
\end{equation}
From the experimental bound on the partial proton lifetime, $\tau (p\to K^+ \nu) > 5.9 \times 10^{33}$ years~\cite{Super-Kamiokande:2014otb}, it follows that $\Gamma(p \to K^{+}\nu)< 3.5 \times 10^{-66}$ GeV, therefore a scenario in which $M_\chi< m_p-m_K^{+}$ would completely rule out our interesting region of the parameter space around the TeV. Considering now the case where $M_\chi > m_p -m_K^{+}$, a natural assumption would be to assume $\chi$ to be Majorana, with a $\frac{1}{2}M_\chi \overline{\chi}_R \chi_R^c$ mass term and to couple it to the SM leptons and the Higgs $H$ via the Yukawa coupling $y_{\nu}$ and provide SM neutrino masses with a standard type-I seesaw. In this case the decay $p \to K^{+} \nu$ would receive contribution from $y_{\nu}$ and one can get an estimate for the decay width
\begin{equation}
\Gamma (p\to K^{+} \nu)
\sim \frac {y_u^2 y_d^2}{64 \pi^3} \frac{m_p^5}{M_\Delta ^4}\Big(\frac{y_{\nu}v}{M_\chi}\Big)^2
\simeq \Big(\frac{y}{10^{-2}}\Big)^{4}\Big(\frac{\text{TeV}}{M_{\Delta}}\Big)^{4}\Big(\frac{\text{TeV}}{M_\chi}\Big)10^{-37}\text{ GeV} \, ,
\label{eq:protonseesaw}
\end{equation}
which again would exclude the most interesting region of our parameter space. Therefore, in order not to spoil the interesting features discussed so far, one needs $M_\chi >m_p-m_K^{+}$ and to disentangle $\chi$ from the generation of neutrino masses, so that the coupling $y_{\nu}$ can be taken as small as needed to respect limits on nucleon's lifetimes, independently of the other parameters.

Moreover, the same couplings in Eq.~\eqref{eq:lagrangiantriplet} which produce the baryon asymmetry in our setup also lead to a partial decay width of $\chi$ given by
\begin{equation} 
\Gamma (\chi \to \overline{udd'})
= \frac{1}{1024 \pi^3}\Big|\frac{y_{u1}y_{d1}}{M_{\Delta_1}^2}+\frac{y_{u2}y_{d2}}{M_{\Delta_2}^2}\Big|^2 M_\chi^5 \,.
\label{eq:washoutpar}\end{equation}
After baryogenesis there is an excess of $\chi$ over $\overline{\chi}$, thus if $\chi$ is Dirac and $M_\chi<M_{\Delta}$ the asymmetry gets erased exactly once $\chi$ decay.
%In fact $\chi\to \overline{udd'}$ and $\overline{\chi}\to udd'$ and these decay will erase the CP asymmetry, due to $\epsilon_d \Gamma_d=-\epsilon_u \Gamma_u$.
Indeed the decays $\chi\to \overline{udd'}$ and $\overline{\chi}\to udd'$ would exactly erase the $CP$ asymmetry, due to $\epsilon_d \Gamma_d=-\epsilon_u \Gamma_u$ following from $CPT$.
Note that $M_\chi\geq M_\Delta$ is not a viable option, because then $\Gamma_u = 0$ and so $\epsilon_d = 0$ via $CPT$, so one would not generate the baryon asymmetry.
Instead, if $\chi$ is Majorana then both $udd$ and $\overline{udd}$ decays are allowed
%and the asymmetry is preserved as long as decays are out of equilibrium when $T=M_\chi$.
and the asymmetry is preserved. However, the asymmetry could be spoiled by inverse decays: any excess of $udd$ or $\overline{udd}$ would go back to $\chi$, that would wash it out upon decaying
Assuming no hierarchies among the Yukawas $y$ we get the following ratio between the rate of inverse decays and Hubble:
\begin{equation}
\frac{\Gamma(\overline{udd'} \to \chi)}{H (T=M_\chi)}
\simeq \frac{(y_u y_d)^2}{256 \pi^3 g^{*^{1/2}}} \frac{\pi}{90^{\frac{1}{2}}} \frac{M_\chi^5}{M_{\Delta}^4}\frac{M_\Pl}{M_\chi^2}
\sim 10^{-7}\Big(\frac{y}{10^{-2}}\Big)^{\!4}\Big(\frac{\text{TeV}}{M_{\Delta}}\Big)^{\!4}\Big(\frac{M_\chi}{\text{GeV}}\Big)^{\!3},
\label{eq:washout_chi}
\end{equation}
%  so we can summarize that the baryon asymmetry in our mechanism does not get erased by $\chi$ washout processes provided that $M_\chi>M_{\Delta}$ with $\chi$ a Dirac fermion or $M_\chi>m_p-m_K$ with $\chi$ a Majorana fermion.
where we have evaluated $H$ at $T=M_\chi$ because that is the range of temperatures for which inverse decays are most efficient: for $T \ll M_\chi$, $\Gamma(\overline{udd'} \to \chi)$ would be exponentially suppressed, for $T\gg M_\chi$, $H$ would be enhanced but not $\Gamma(\overline{udd'} \to \chi)$.

To summarize, $\chi$ does not alter our results as long as i) to avoid nucleons decays, $M_\chi >m_p-m_K^{+}$ and its coupling to neutrinos is small enough (so that it cannot play a role in the generation of neutrino masses) and ii) to avoid further washouts, $M_\chi$ is Majorana and it is smaller enough than $M_\Delta$ to keep $\Gamma(\overline{udd'} \to \chi)/H \ll 1$, see Eq.~(\ref{eq:washout_chi}).

\subsection{Signals of the composite scalar $\Delta$ in laboratory experiments}
\subsubsection{Collider searches} \label{sec:collider}

As it turns out (see Fig.~\ref{fig:parspace}), one can easily accomodate baryogenesis from supercooled confinement with the scalar $\Delta$ around or below the electroweak scale, the properties of these particles can be probed thanks to their signatures in high-energy colliders.

A first lower bound on the mass of $\Delta$ comes from the measurement of the decay width of the $Z$ boson performed at LEP~\cite{ALEPH:2005ab}. Since $\Delta$ interacts weakly due to its $U(1)_Y$ charge, it can contribute to the $Z$ boson width, provided that $M_\Delta<M_Z/2$. We compute the following expression for the decay width of $Z$ into a $\Delta \Delta^*$ pair
\begin{equation}
\Gamma (Z\to \Delta \Delta^*)=\frac{3}{16\pi}\Big(\frac{2g \sin^2 \theta_W}{3\cos \theta_W}\Big)^2\,M_Z \Big(1-\frac{4 M_\Delta^2}{M_Z^2}\Big)^{3/2}\, .
\end{equation}
Since this value is orders of magnitude larger than the uncertainty on the $Z$ boson width, we exclude $M_{\Delta}<45$ GeV.
%Since this value is about one order of magnitude larger than the uncertainty on the $Z$ boson width measured by CMS~\cite{CMS:2022ett}, $M_{\Delta}<45$ GeV is thus excluded. \FS{CMS or LEP?}\vdr{The LEP reference is about the total decay width, we are interested in an updated measurement of the invisible one to exclude $M_\Delta<M_Z/2$, so I would keep only the CMS reference}

Then one can note that $\Delta$ has the same quantum numbers of a right up-type squark in SUSY, therefore it will share its same production mechanisms at colliders. To our knowledge, the closest searches to our scenario that have been performed  are those involving squark decays into diquarks, through R-parity violating (RPV) couplings.
Constraints on top squarks decaying through RPV couplings were first set by the ALEPH experiment at LEP~\cite{ALEPH:2002nmz}, excluding at 95\% CL masses below 75 GeV via four-jet searches. More stringent bounds come from further searches performed by CDF at Tevatron~\cite{CDF:2013yum} and by ATLAS~\cite{ATLAS:2017jnp} and CMS~\cite{CMS:2022usq} at the LHC. The combination of these searches rules out top squarks with RPV decay in the $45 \text{ GeV}<M_\Delta<770$ GeV interval. All these searches aim to detect four jets in the final state associated to the decay of pair produced heavy resonances. This is the typical signature that one would expect associated with $\Delta$, as for the allowed values of its mass and coupling constants it would decay promptly in the detector. 
However, as the diquark and leptoquark couplings of $\Delta$ are expected to be of the same size, some significant branching ratio of $\Delta$ scalar could reside 
in the $u\,\chi$ channel, such that the signal would be different than the one assumed in four-jet searches, and the limits would change.
%Translating these searches to precise bounds on the $\Delta$ parameter space is non-trivial and beyond the scope of our work, but for the purpose of our analysis we included the above mentioned constraints in the parameter space in  Fig.~\ref{fig:parspace}. 
Performing a detailed collider study of the $\Delta$'s, depending on their branching rations into two-jet and one-jets plus missing energy, is beyond the scope of our work, so for definiteness we use the constraints that assume $\Delta$ to dominantly decay into two jets.

\subsubsection{Neutron electric dipole moment}
As we have seen in the previous section, the new fields $\Delta_i$ and $\chi$ contain new sources of $CP$ violation and therefore give rise to new contributions to the neutron electric dipole moment (nEDM). The experimental bound set by the nEDM collaboration at the Paul Scherrer Institut is given by $|d_n|<1.8\times\, 10^{-26} e\cdot cm$~\cite{Abel:2020pzs}. In this section we analyze the impact of this bound on our model and show that it does not place further significant constraints on the parameter space.

In an EFT approach the electromagnetic moments of a fermion $\psi$ can be described by the effective Lagrangian
\begin{equation}
\mathcal{L}^{\text{eff}}=c_R\overline{\psi}\sigma_{\mu\nu}P_R \psi F^{\mu\nu} + \hc,
\qquad \text{where} \qquad 
d=i (c_R-c_R^{*})=-2 \IM\,c_R 
\end{equation}
is the EDM of $\psi$ and $F_{\mu\nu}$ is the field strength tensor of the photon.
The diagrams which may give a non-vanishing contribution to the %$CP$ odd \FS{all EDMs are CP-odd, if it is not }
nEDM must have an imaginary component.
In general the amplitude of any loop diagram can be recasted as a sum of amplitudes each coming by operators $O_i$ weighted by their coefficients $C_i$, which can be written as $C= y\,F$. 
In this notation $y$ stands for the product of the couplings of all the vertices and $F$ is a function coming from the loop integration~\cite {Bensalem:2021qtj}. In principle both $y$ and $F$ are allowed to be complex, but diagrams with the heavy $\Delta$ and $\chi$ running in the loop cannot have all the intermediate states go on-shell. Therefore the optical theorem implies that $\IM (F)=0$ and only $y$ can be complex.
Moreover, any combination of couplings entering the expression of the nEDM should be invariant under rephasing of the fields, thus we need at least 4 couplings involving the new BSM states, because $Y=y_{d1}^{*}y_{u1}y_{u2}^{*}y_{d2}$ is the simplest combination of the Yukawas with at least one physical phase. As a consequence one needs to go at least to 2-loop to find a possible non-zero contribution. 

\begin{figure}[!hbt]
 \centering
  \includegraphics[width=0.9\linewidth]{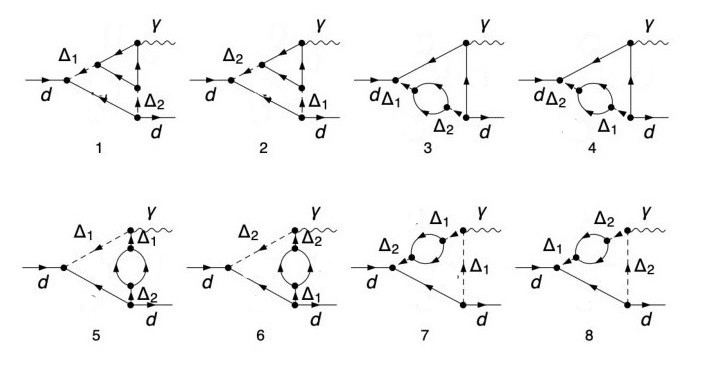}
\caption{Relevant topologies at two loops for the nEDM. Several contributions for each
topology occur due to flavour. While diagrams in the upper row are manifestly symmetric
under $\Delta_1 \leftrightarrow  \Delta_2$ exchange, a more careful analysis is needed to show the cancellation among
the diagrams in the lower row, which is non-trivial due to the different number of $\Delta_i$ propagators involved when the photon is attached to an internal $\Delta$ leg.}\label{fig:Deltanedm}
\end{figure}

The relevant topologies of our model are depicted in Fig.~\ref{fig:Deltanedm}, %they are so called Barr-Zee like diagrams.
they include the Barr-Zee diagrams~\cite{Barr:1990vd}.
They all give in principle non-zero contributions to the nEDM. However, for each of them one can always take the contribution of
%\vdr{One or more (a set), "mirror diagram", "time-reversed diagram"...this has to be clarified}
the `mirrored' diagram, with complex-conjugate Yukawa couplings and with the photon attached to the same type of field. One can decompose the sum of each set of conjugate diagrams as 
\begin{equation}
\sum_{j=1}^{\#\text{conj.}} \mathcal{A}_j= Y F( M_{\Delta_1}, M_{\Delta_2}) +Y^{*}  F( M_{\Delta_2}, M_{\Delta_1})\, ,
\label{eq:sumEDMs}
\end{equation}
where $F(i,j)$ is coming from the loop integration. $F(i,j)$ is not generically a symmetric function of the scalar masses, but the asymmetry scales like $ F (M_{\Delta_2}, M_{\Delta_1})=F(M_{\Delta_1}, M_{\Delta_2})(1+\mathcal{O}(k^{2}/M_{\Delta_{1,2}}^2))$, therefore it vanishes in the $k^2=0$ limit for which the nEDM is measured. An explicit proof of this statement is given in the Appendix~\ref{sec:app1}. Therefore the sum %of the conjugate diagrams
in Eq.~(\ref{eq:sumEDMs}) ends up to be real and thus 2-loop diagrams do not produce a non-vanishing nEDM.
%There may be contributions to the nEDM at the three loop level, which we estimate to be of the order of \vdr{Insert an estimate for the 3 loop}. Therefore, the constraint set by nEDM on the mass of $\Delta$ turns out to be much less stringent than those coming from colliders. 

At the 3-loop level one expects the dominant contributions to the EDMs to scale as
\begin{equation}
d \sim \frac{y^2 g^4}{(16\pi^2)^3} \times \text{CKM} \times \frac{m_t}{M_\Delta^2}
\simeq \frac{10^{-11}}{\text{TeV}} \Big(\frac{y}{0.1}\Big)^2 \Big(\frac{\text{TeV}}{M_\Delta}\Big)^2 \frac{\text{CKM}}{0.1} \,,
\label{eq:EDM3loops}
\end{equation}
where $m_t$ is the top-quark mass, $g$ is the weak gauge coupling and by `CKM' we mean a combination of non-diagonal elements of the CKM matrix, which is unavoidable to ensure $CP$ violation and is smaller than at least 0.1. Since limits on coefficients of quarks EDMs (and also CEDMs, for which the discussion is analogous) are in the ballpark of $10^{-9}/$TeV (see e.g.~\cite{Sala:2013osa}) and $M_\Delta \geq \text{TeV}$ from collider bounds, then 3-loop contributions do not give rise to observable EDMs. To conclude, the constraint set by nucleon EDMs on the mass of $\Delta$ turns out to be much less stringent than those coming from colliders.

\subsubsection{Flavour violation}
Flavour-changing neutral current (FCNC) observables are expected to set powerful constraints on BSM models, due to the non trivial flavour structure that they imply in order to comply with experimental data. In our scenario, $\Delta F=2$ processes are prohibited at tree level due to the antisymmetric couplings of $\Delta$ with the diquarks, which can be parametrized as $y_{ij}=\epsilon_{ijk}y_k$, where $i,j,k$ are flavour indices.
These bounds were firstly discussed in~\cite{Giudice:2011ak} and one can identify our state $\Delta$ with the diquark denoted as $\RN{8}$ there. 

First, it has to be pointed out that the most general flavour structure, with all the flavour off-diagonal couplings switched on, is not needed to account for baryogenesis, as a single non-zero complex coupling is sufficient to provide a physical CP violating phase. In this setup flavour bounds would be evaded completely.

\begin{figure}[!hbt] 
\centering
  \includegraphics[height=3cm, width=11cm]{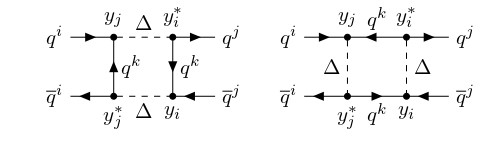}
\caption{ Loop level exchange of $\Delta$ contributing to $\Delta F=2$ transitions.}\label{fig:Deltaf2}
\end{figure}

If one wants to stick to the general scenario, bounds coming from the mixing of the neutral $K$ and $B_{d,s}$ mesons have to be taken into account.
The $\Delta$ scalars then enter the box diagrams in Fig.~\ref{fig:Deltaf2} and they generate a four-fermion effective Hamiltonian given by
\begin{equation} \label{eq:deltaf}
H_{eff}^{NP}(\Delta F=2)=\frac{1}{32 \pi^2}\biggl(\frac{\lambda_i \lambda_j^*}{M_{\Delta}^2}\biggl)G\Big(\frac{m_k^2}{M_{\Delta}^2}\Big)\big(\overline{q_i}\,\gamma^{\mu}P_Rq_j\big)\big(\overline{q_i}\gamma_{\mu}P_R\,q_j\big) \,,
\end{equation}
where $i$ and $j$ are the external quark flavour indices, $k\neq i\neq j$ is the flavour index of the internal quark, colour indices are contracted within the brackets and the loop function $G$ is given by~\cite{Giudice:2011ak}
\begin{equation}G(x)=\frac{1-x^2+2\,x \text{ log }x}{(1-x)^3} \,.
\end{equation}
Bounds on representative $D=6$ effective operators contributing to $\Delta F=2$ transitions have recently been obtained by the UTfit collaboration~\cite{Bona:2022zhn}, from a combination of CP conserving and CP violating observables. We can apply these bounds to our model, results are summarised in Table~\ref{tab:flavour}.
Other constraints which we include in Table~\ref{tab:flavour} are those associated to the $\Delta F=1$ processes $b\to s \gamma$, $b\to d \gamma$, $R_b$ and $B^{\pm}\to \phi \pi^{\pm}$. We refer to~\cite{Giudice:2011ak} for more details as the constraints reported there can be easily translated to our notation.

One can note that  even the most constraining observable, namely $\epsilon_k$, is compatible with $\Delta$ diquark couplings of $\mathcal{O}(0.01)$ around the TeV scale, which, in the absence of major hierarchies in the Yukawa couplings, is the benchmark for our baryogenesis mechanism to work. Therefore, we don't expect flavour physics to exclude values of $M_{\Delta}$ that are not already excluded by collider searches.  In fact, even assuming a hierarchy of the type $y_{ui} \gg y_{di}$, in order to lower the bound on $M_{\Delta}$ from colliders, one would get at the same time weaker bounds from flavour. 

\begin{table}[tb]
\centering
\begin{tabular}{ |c c |}
\toprule
Observable & Bound/($M_\Delta$/TeV) \\ 
\hline
$\epsilon_K$ & $|\IM(y_1 y_2^{*})|\leq7.8\times 10^{-4}$ \\
$\Delta m_K$ & $|\RE(y_1 y_2^{*})|\leq 1.4\times 10^{-2}$ \\
$B_d$ mixing & $|y_1 y_3^{*}|\leq 1.7\times 10^{-2}$ \\
$B_s$ mixing & $|y_2 y_3^{*}|\leq5.8\times 10^{-2}$ \\
$b\to s \gamma$ & $|y_2 y_3^*|\leq2.5$\\
$b\to d \gamma$ &  $|y_1 y_3^*|\leq 1.2$ \\
$R_b$ &  $|y_{1,2}|\leq 24$ \\
$B^{\pm}\to \phi \pi^{\pm}$ &  $|y_1 y_3^*|\leq 3.6\times 10^{-3}$\\
\bottomrule
\end{tabular}
\caption{Bounds in units of $M_{\Delta}$/TeV on antisymmetrically coupled diquarks.
\label{tab:flavour}}
\end{table}

To conclude, the scalar realisation of our baryogenesis framework may well show-up in current and near-future flavour experiments, but the non-observation of flavour or $CP$-violation there would only exclude some very specific choices of its parameters and not preclude the realisation of baryogenesis.

  \begin{figure}
\centering
  \begin{minipage}[b]{0.48\textwidth}
   \includegraphics[height= 7cm, width=8cm]{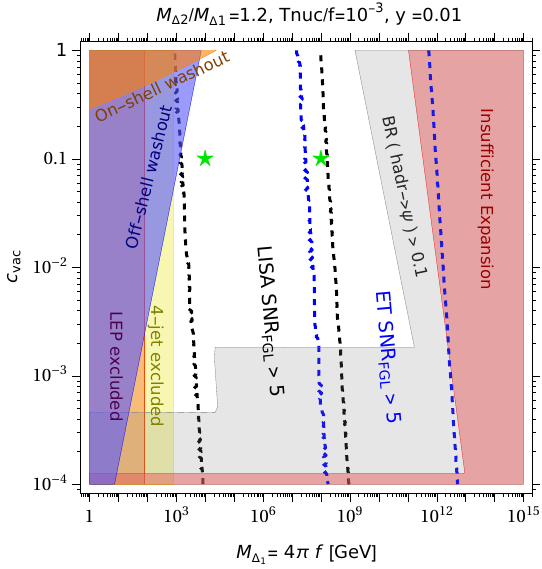}
    \end{minipage}
    \vspace{0.5cm}
   \hfill
 \begin{minipage}[b]{0.48\textwidth}
\includegraphics[height= 7cm, width=8cm]{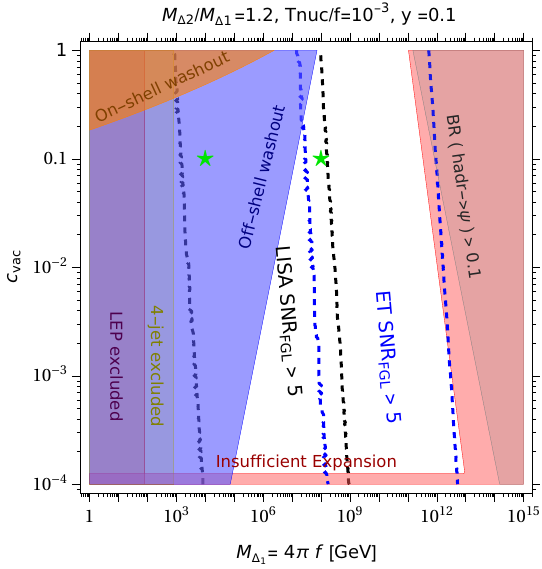}
  \end{minipage}  
  \vspace{0.5cm}
  \begin{minipage}[b]{0.48\textwidth}
   \includegraphics[height= 7cm, width=8cm]{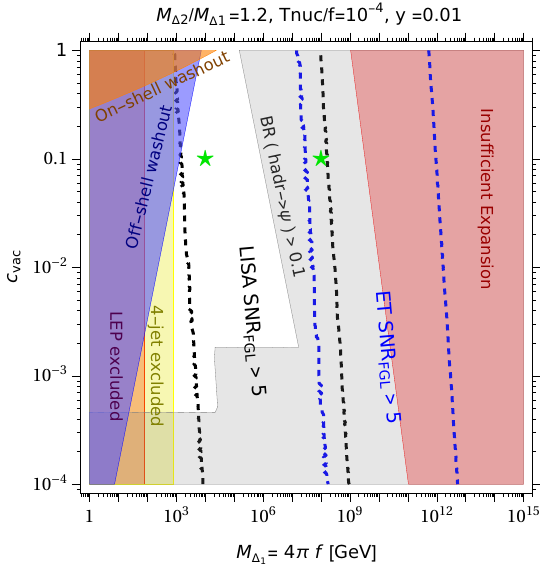}
    \end{minipage}
    \hfill
 \begin{minipage}[b]{0.48\textwidth}
\includegraphics[height= 7cm, width=8cm]{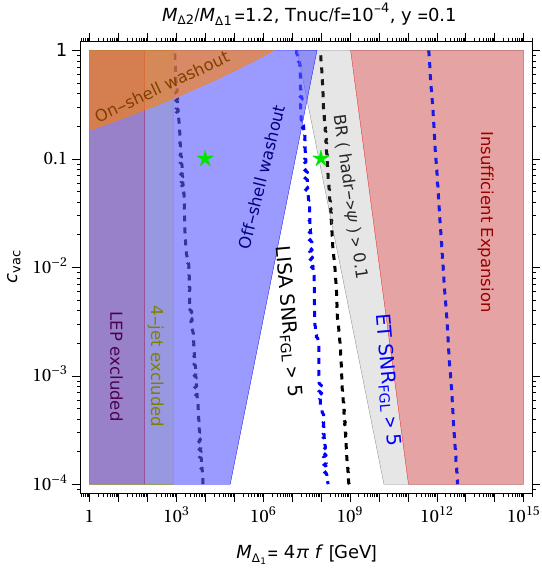}
  \end{minipage}
 
  \caption{Parameter space for the composite scalar realisation of our framework, which realises baryogenesis.
  We choose $g_\TC = 80$, $g_* = g_\SM + g_\TC$, $g_\RH = g_\SM + 1$ and $Z=1$ for definiteness.
  The white area is allowed and accommodates the observed baryon asymmetry, the red and gray regions are excluded by the PT dynamics, as in Fig.~\ref{fig:parspace_framework}.
  We find that $\Delta$ annihilations are slower than $\Delta$ decays in the entire parameter space we plot, hence our computation of the BAU is always valid.
   A value of $y$ smaller than unity is needed to avoid washout. Off-shell and on-shell washout regions are shaded to exclude parameter space where, respectively, $2\to2$ interactions Eq.~\eqref{eq:offshell} and inverse decays Eq.~\eqref{eq:onshell} are in equilibrium.
  Constraints from collider experiments include LEP searches,  excluding values of $M_{\Delta}<75$ GeV and four-jet searches performed at Tevatron and at the LHC, excluding $\Delta$ in the $50 \text{ GeV}<M_{\Delta}<770 \text{ GeV}$ range.
  We also show the parameter space testable by the Einstein Telescope (ET) and LISA using  $\alpha_\GW=100$, $\beta/H=15$, where the astrophysical foregrounds have been taken into account. We denote with red stars the two points of the parameter space for which we computed the gravitational wave spectrum shown in Fig.~\ref{fig:gwspectrum}.
}
\label{fig:parspace}
\end{figure}

\subsection{Summary of the results}

We summarize our results in the plots of Fig.~\ref{fig:parspace}. We have chosen two different benchmark values for $y=\{0.01,0.1\}$ and $\Tnuc/f=\{10^{-3},10^{-4}\}$, $M_{\Delta_1}=4\pi f$ and the dilaton wavefunction normalization $Z=1$. We compute at each point the boost factor at bubbles collision $\gcoll$ using Eq.~\eqref{eq:gammawp} and we set $M_{\Delta_2}=1.2\,M_{\Delta_1}$, $g_\TC=80$, $g_\RH=107.75$ ($=g_\SM + 1$, where the +1 accounts for the dilaton) and $g_*= g_\TC + g_\SM$.
We saturate the baryon yield at each point, $Y_B/Y_B^{\text{obs}}=1$, by obtaining $ Br\big(\text{hadr}\to \Delta\big)$ from Eq.~\eqref{eq:Yield}.
As anticipated in Sec.~\ref{sec:framework} we see that $Y_B$ gets suppressed in the runaway regime, hence the allowed parameter space closes at large $f$ and/or small $\Tnuc/f$.
Values of $f$ lower than the weak scale, for large $\cvac$, cannot be reached due to washout effects. $\Delta$ annihilations before decays instead turn out to not constrain our parameter space, for the values of the parameters that we have chosen.

We also display in Fig.~\ref{fig:parspace} collider bounds, discussed in Sec.~\ref{sec:collider}, and the expected LISA and ET sensitivities to the GW signal from the PT, discussed in Sec.~\ref{sec:GWs}, where to avoid clutter we display only those that include astrophysical foregrounds.
%Collider bounds, discussed in \ref{sec:collider},  and the expected gravitational wave signals are also shown.  
By comparing our plots with Fig.~6 of~\cite{Baldes:2021vyz}, we see that our scenario extends the allowed parameter space by many order of magnitudes towards smaller values of $M_{\Delta}$, allowing the PT to occur down to the electroweak scale, potentially in reach of LISA and the Einstein Telescope, and making the mechanism testable at the LHC.

\section{Model 2: Leptogenesis from a composite sterile neutrino}
\label{sec:leptogenesis}

\subsection{Model, neutrino masses and the baryon asymmetry}
We now apply our framework for the generation of the baryon asymmetry, based on a supercooled confining PT, to a different model where the observed BAU is sourced via leptogenesis~\cite{Fukugita:1986hr} through the decays of composite sterile neutrinos, which in addition provide mass to the SM neutrinos via a seesaw mechanism.
This way we both substantiate the statement that our framework is more general than its specific realisations, and we find results that are interesting for leptogenesis per-se.
%We would like to show explicitly that the framework for the generation of the baryon asymmetry discussed so far, based on a supercooled PT involving a confining sector, is indeed a general paradigm not limited to a specific realization, by applying it to a different model, where the observed BAU is sourced via leptogenesis~\cite{Fukugita:1986hr} through the decays of sterile fermions, which in addition  provide mass to the SM neutrinos via a seesaw mechanism.\\  

%We introduce the sterile neutrinos $N_a $ as hadrons of the confining sector driving the PT, hence we assume the parametric relation $M_{N_a}\sim \mathcal{O}(\pi f)$ to hold.
We introduce the sterile neutrinos $N_a$ (and denote their two two-component spinors as $N_{aL}$ and $N_{aR}$) as hadrons of the confining sector, that is responsible for their Dirac mass $M_{N_a}\sim \mathcal{O}(\pi f)$.
%We choose the inverse seesaw~\cite{Mohapatra:1986aw,Mohapatra:1986bd} (ISS) as our setup to produce the observed BAU since it can generate the observed neutrino masses via new physics at an experimentally accessible scale. The ISS Lagrangian is given by 
We then assume that a portal between the confining sector and the SM breaks the SM lepton number, generating inverse seesaw~\cite{Mohapatra:1986aw,Mohapatra:1986bd} (ISS) as 
\begin{equation}
-\mathcal{L}^{ISS}= y_{a \alpha}\overline{N}_{aR} H L_{\alpha}+ M_{N_a}\overline{N}_a N_a+\frac{\mu_{ab}}{2}\overline{N}_{aL}^c\,N_{bL} +\hc ,
\end{equation}
where $L$ are the left-handed SM leptons, $\alpha=\{ e, \mu, \tau\}$, $y$ is the Yukawa coupling matrix between active and sterile neutrinos, $\mu$ is a lepton number breaking Majorana mass matrix, where we further denote by $\mu_a$ ($\overline{\mu}$) its complex diagonal (off-diagonal) entries.
The presence of $\mu$ induces at low energies the $d=5$ Weinberg operator~\cite{Weinberg:1979sa}
\begin{equation} \label{eq:Weinberg}
\delta \mathcal{L}^{d=5}=c_{\alpha\beta}^{d=5}\Big(\overline{L^c_\alpha}\, \tilde{H}^{*}\Big)\Big(\tilde{H}^{\dagger}L_\beta\Big)\,,
\end{equation}
where $\tilde{H}=i \sigma_2H^{*}$ and 
\begin{equation} \label{eq:ISSWeinberg}
c_{\alpha\beta}^{d=5}= \Big(y^T \frac{1}{M_{N}^T}\mu\frac{1}{M_N}y\Big)_{\alpha\beta}.
\end{equation}
%
%We assume the heavy Dirac states $N_a$ to come in two generations, which is the minimum number needed to comply with neutrino oscillations data. Without loss of generality, we choose a basis where $M_N$ is diagonal and real.
Upon electroweak symmetry breaking, the mass matrix $m_\nu$ for the light neutrinos is given by  
\begin{equation} \label{eq:ISSneutrino}
m_{\nu}= \frac{v^2}{2} y^T M_N^{T -1} \mu M_N^{-1} y \, ,
\end{equation}
one can thus express $\mu=\mu(f)$ for a fixed value of the Yukawa coupling $y$, assuming $m_{\nu}$ to be of the order of the atmospheric neutrino mass, $m_{\nu}\sim 0.05$ eV.

We now assume the heavy Dirac states $N_a$ to come in two generations, which is the minimum number needed to comply with neutrino oscillations data. Without loss of generality, we choose a basis where $M_N$ is diagonal and real.
In the non confining ISS the parameter $\mu$ is technically small but the hierarchy $\mu \ll M_N$ is not set by any concrete theoretical justification. The latter is instead naturally justified in our setup of composite sterile fermions.
One can indeed envisage the possibility that, while the Dirac mass $M_{N}$ originates from the confining sector, the Majorana one $\mu$ originates from physics external to the confining sector, offering a natural way to accomodate $\mu \ll M_N$.
An explicit realization is provided in~\cite{Chacko:2020zze}, where the strong dynamics is that of a conformal sector deformed by a relevant operator so that it confines in the IR (giving rise to the Dirac masses $M_N$), and by another operator with different scaling dimension and a small coefficient that gives rise to $\mu \ll M_N$.
%Another possibility is to assume that an approximately UV-conformal sector confines in the IR due to obtain the hierarchy from the scaling dimensions of operators in the approximately UV-conformal sector that confines at the scale $f$, as shown in~\cite{Chacko:2020zze}.
As anticipated in Sec.~\ref{sec:framework}, our framework also leads to the expectations $\Delta M \equiv |M_{N_2} - M_{N_1}| \ll M_{N_1}$, because the $N$'s are all hadrons of the same sector, as well as $y_{a\alpha} \ll 1$, because the Yukawas originate from a portal.

%While in the non confining ISS the parameter $\mu$ is technically small but its value is not set by any fundamental scale, $\mu \ll f$ can be instead naturally obtained in our setup of composite sterile fermions.
%For instance, if the hidden sector is approximately conformal in the UV, one can obtain a value of $\mu$ parametrically smaller than the compositeness scale $f$, arising from the scaling dimensions of operators in the CFT, as shown in ~\cite{Chacko:2020zze}.
%Therefore, under the natural assumption that $\mu_a \sim \overline{\mu}\sim\mu \ll M_{N_a}, \Delta_M$, where $\Delta_M=|M_{N_2}-M_{N_1}|$, one can diagonalize the sterile neutrinos mass matrix to first order in $\delta_a=\mu/M_{N_a}$.
Therefore, under the natural assumption that $\mu_a \sim \overline{\mu}\sim\mu \ll M_{N_{1,2}}, \Delta M$, one can diagonalize the sterile neutrinos mass matrix to first order in $\delta_a=\mu/M_{N_a}$.
Defining four Majorana states $\tilde{N}_i,$ $i=1,2,3,4$, with masses $M_i$, we have
\begin{equation} 
-\mathcal{L}^{ISS}_{\text{mass}} \supset h_{i\alpha}\overline{\tilde{N}}_i\,H L_{\alpha}+\frac{1}{2}M_i\overline{\tilde{N}}_i\tilde{N}_i +\hc ,
\end{equation}
To first order in $\delta_a$ their masses and couplings are given by~\cite{Agashe:2018cuf}
\begin{equation}
\begin{split} 
\label{eq:ISSmass}
M_1 \simeq M_{N_1}\Big(1-\frac{\delta_1}{2}\Big), \hspace{1cm} h_{1\alpha}\simeq \frac{i}{\sqrt{2}}\Big(y_{1\alpha}+\frac{\delta_1}{4}y_{1\alpha}+\overline{\delta}_1y_{2\alpha}\Big) \\
M_2 \simeq M_{N_1}\Big(1+\frac{\delta_1}{2}\Big) , \hspace{1cm} h_{2\alpha}\simeq \frac{1}{\sqrt{2}}\Big(y_{1\alpha}-\frac{\delta_1}{4}y_{1\alpha}-\overline{\delta}_1y_{2\alpha}\Big) \\
M_3 \simeq M_{N_2}\Big(1-\frac{\delta_2}{2}\Big) , \hspace{1cm} h_{3\alpha}\simeq \frac{i}{\sqrt{2}}\Big(y_{2\alpha}+\frac{\delta_2}{4}y_{2\alpha}-\overline{\delta}_2y_{1\alpha}\Big) \\
M_4 \simeq M_{N_2}\Big(1+\frac{\delta_2}{2}\Big) , \hspace{1cm} h_{4\alpha}\simeq \frac{1}{\sqrt{2}}\Big(y_{2\alpha}-\frac{\delta_2}{4}y_{2\alpha}+\overline{\delta}_2y_{1\alpha}\Big) 
\end{split}
\end{equation}
where we defined 
\begin{equation}
\overline{\delta}_i=\frac{\overline{\mu} M_{N_i}}{M_{N_2}^2-M_{N_1}^2}\,. \label{eq:degeneracy}
\end{equation}
Therefore we can see from Eq.~\eqref{eq:ISSmass} that the heavy neutrinos form pseudo-Dirac pairs with small Majorana splitting controlled by $\mu/M_{N_i}$.

The $CP$ asymmetry generated by the decays of the sterile neutrino $N_a$ %is defined as
then reads
\begin{equation} 
\epsilon_a
\equiv \frac{\sum_\alpha \Big[\Gamma\Big(N_a \to \ell_\alpha H\Big)-\Gamma\Big(N_a \to \overline{\ell_\alpha} H^{*}\Big)\Big]}{\sum_\alpha \Big[\Gamma\Big(N_a \to \ell_\alpha H\Big)+\Gamma\Big(N_a \to \overline{\ell_\alpha} H^{*}\Big)\Big]}
= \frac{1}{8\pi}\sum_{j \neq i}\frac{\IM[(h h^\dagger)^2_{ij}}{(h h^\dagger)_{ii}}f_{ij}\, , 
 \label{eq:ISSleptogenesis}
\end{equation}
where we summed over the lepton flavour $\alpha$ in the final state and $f_{ij}\equiv f_{ij}^V + f_{ij}^{\text{self}}$ includes contributions given by vertex and self-energy corrections to the decay, which are given by~\cite{Covi:1996wh}
\begin{equation}
f_{ij}^{V}=g\Big(\frac{m_j^2}{m_i^2}\Big), \hspace{1cm}  g(x)=\sqrt{x}\Big[1-(1+x)\,\text{log }\frac{x+1}{x}\Big]
\end{equation}
and~\cite{Deppisch:2010fr}
\begin{equation}
f_{ij}^{\text{self}}=\frac{(m_i^2-m_j^2)m_im_j}{(m_i^2-m_j^2)^2+m_i^2 \Gamma_j^2},
\end{equation}
where $\Gamma_{i}\sim(y y^\dag)_{ii} M_i/16\pi$ is the decay width of $N_i$ and we assume no hierarchies in masses or couplings among the singlet generations and the SM flavours.

\begin{figure}[tb!]
  \centering
  \includegraphics[width=0.6\linewidth]{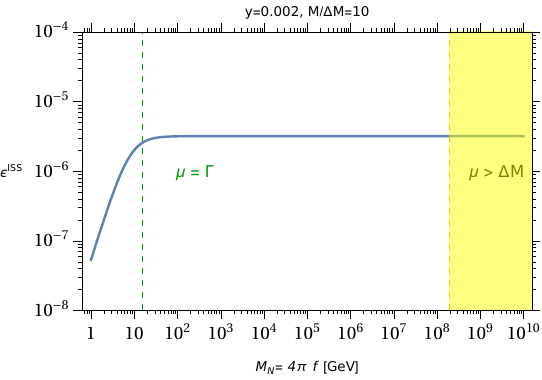}
  \caption{Rate asymmetry $\epsilon^{ISS}$ in the inverse see-saw mechanism, obtained from Eq.~\eqref{eq:ISSasymmetry}, for a fixed reference value of the Yukawa coupling $y=0.002$, a degeneracy $M/\Delta M =10$ and $\mu=\mu(f)$ fixed by neutrino masses. One can note that an asymptotic value for the produced asymmetry is reached in the $\mu \geq \Gamma$ regime. %We have excluded the region on the right for which $\mu >\Delta_M$ and Eq.~\eqref{eq:ISSasymmetry} breaks down, which translates into an upper bound of $M_N$.
  In the yellow region at large $M_N$ our derivation of Eq.~\eqref{eq:ISSasymmetry} breaks down, because it assumed $\mu  \ll \Delta M$.}
    \label{fig:epsilonISS}
\end{figure}

Let us now focus on the $\epsilon_1$ and $\epsilon_2$ contributions to Eq.~\eqref{eq:ISSleptogenesis}:
\begin{equation} \begin{split} 
\epsilon_1 &= \frac{1}{8\pi (h h^\dagger)_{11}} \IM\Big[(h h^\dagger)^2_{12}f_{12}+(h h^\dagger)^2_{13}f_{13}+(h h^\dagger)^2_{14}f_{14}\Big]\,, \\
\epsilon_2 &= \frac{1}{8\pi (h h^\dagger)_{22}} \IM\Big[(h h^\dagger)^2_{21}f_{21}+(h h^\dagger)^2_{23}f_{23}+(h h^\dagger)^2_{24}f_{24}\Big] \,.
\end{split}\end{equation}
Due to the pseudo-Dirac nature of the ($\tilde{N}_1, \tilde{N}_2$) pair, it follows that
\begin{equation}
(h h^\dagger)^2_{13}\simeq -(h h^\dagger)^2_{23}\simeq -(h h^\dagger)^2_{14}\simeq (h h^\dagger)^2_{24}; \hspace{1cm} f_{13}\simeq f_{14}\simeq f_{23}\simeq f_{24}.
\end{equation}
As a consequence, when considering $\epsilon_1+\epsilon_2$, the $f_{12}$ and $f_{21}$ terms will dominate the sum as those involving $f_{i3}$ and $f_{i4}$ turn out to be of higher order in $\delta$ and $\overline{\delta}$ when summed up. An analytical parametric expression for $\epsilon$ in the ISS was obtained by the authors of~\cite{Agashe:2018cuf}, $\epsilon \sim \mu ^2/(M_N\,\Gamma)$, where we dropped the family indices for $\mu$ and $\Gamma$ to show only the parametric dependence.
However, the above expression is assuming $\mu \ll \Gamma$, which is not strictly required by the ISS mechanism. 
%A more general relation can be derived assuming no hierarchy among $\mu$ and $\Gamma$.
We then derive a more general relation that assumes no hierarchy among $\mu$ and $\Gamma$.
In fact as long as $\mu \ll M_{N}$, $\Delta M$ and $\Gamma \ll M_{N}$, $\Delta M$ are satisfied, the dominant one-loop contribution to the rate asymmetry is given by the self energy correction of $N$, because one has $f_{12}^V-f_{21}^{V} \sim \delta_1$, therefore the terms involving $f_{12}^{\text{self}}$ and $f_{21}^{\text{self}}\simeq-f_{12}^{\text{self}}$ dominate in $\epsilon_1+\epsilon_2$.
We then obtain the total rate asymmetry $\epsilon^{ISS}$ 
\begin{equation}
\epsilon^{ISS}
\simeq 2(\epsilon_1+\epsilon_2)
\simeq \frac{\IM(y y^\dagger)_{12}}{\pi}\,\overline{\delta}_1 \,f_{12}^{\text{self}}
\simeq \frac{\text{Im}y^2}{\pi} \Big(\frac{\mu}{M_N}\Big)^{\!2} \frac{M_N}{\Delta M} \frac{M_N^2}{4\mu^2 + \Gamma^2}
%+\mathcal{O}(\delta^2, \delta\, \overline{\delta}, \overline{\delta}^2)
\label{eq:ISSasymmetry}\, ,
\end{equation}
%
%which we plot for clarity as a function of $M_N\sim M_{N_{1,2}}$ in Fig.~\ref{fig:epsilonISS}.
where the factor of 2 includes the contribution from $\epsilon_3+\epsilon_4$, which has the same parametric dependence as the one from $\epsilon_1+\epsilon_2$.
We plot $\epsilon^{ISS}$ as a function of $M_N\sim M_{N_{1,2}}$ in Fig.~\ref{fig:epsilonISS}.
One can observe from Eq.~\eqref{eq:ISSasymmetry} that an enhancement of the rate can be obtained in the presence of a quasi-degeneracy among the Dirac masses $M_{N_a}$. When the two generations $a=1,2$ are nearly degenerate, the $CP$ asymmetry is enhanced compared to the non-degenerate case by a factor $M_N/2 \Delta M$, as follows from Eq.~\eqref{eq:degeneracy}.
We stress that Eq.~(\ref{eq:ISSasymmetry}) holds for $\mu  \ll \Delta M$, hence we exclude regions of parameter space where this does not hold from our study. A generalization to the case $\mu  \gtrsim \Delta M$ is beyond the scope of this work.

As the baryon asymmetry is now sourced by a lepton asymmetry, one has to require that the reheating temperature, Eq.~\eqref{eq:TRH}, is larger than the sphalerons decoupling temperature, $T_\RH \geq T_{\text{sph}}\simeq 132$ GeV, in order for the sphalerons to be in thermal equilibrium and therefore to transfer the lepton asymmetry to the baryon sector.
Finally, the baryon asymmetry is obtained by plugging $\epsilon^{ISS}$ of Eq.~(\ref{eq:ISSasymmetry}) in $Y_B$ of Eq.~(\ref{eq:Yield}).

\subsection{Washout processes}

Following analogous steps to those discussed in the model with the $\Delta$ scalar, we now have to take into account washout effects of the lepton number produced at $T_\RH$. The $2\to2$ scattering rate $\ell H\leftrightarrow  \overline{\ell} H^{*}$, mediated by the sterile fermion $N$, violates lepton number. 
Therefore its rate must vanish in the $\mu \to 0$ limit and it can be estimated from the coefficient of the Weinberg operator in the ISS, Eq.~\eqref{eq:ISSWeinberg}:
\begin{equation}
\Gamma_{WO}(\ell H\to \overline{\ell}H^*)\simeq \frac{y^4 T^3_\RH \mu^2}{4\pi M_N^4}\,,  \label{eq:ISSwashout}
\end{equation} 
which presents a $\mu^2/M_N^2$ suppression with respect to the standard Type-I Seesaw one, where one has only one scale $M_N \sim \mu$.
We also included a factor of 2 in Eq.~\eqref{eq:ISSwashout} to account for the contributions given by the degenerate neutrinos. Moreover we assume these washouts to be efficient for $T_\RH \geq m_H/3$, when a sizable population of $H$ is still present. Anyway the precise value of $T_\RH$ for the decoupling of washout effects is not crucial as these processes turn out to play a negligible role in our parameter space. 
In fact this rate does not spoil the lepton number asymmetry produced by the decays of $N$ provided that it is smaller than $H(T_\RH)$, %so one obtains
i.e. if
\begin{equation}
T_\RH< \sqrt{\frac{\pi^2}{90}}
\frac{\pi v^4\,g^{*^{1/2}}}{M_\Pl\,m_{\nu}^2}
\sim 10^{13}\text{GeV}\, \label{eq:ISSoffshell} \,,
\end{equation}
where the dependence of $\Gamma_{WO}$ on $y$, $\mu$ and $M_N$  has been absorbed into $m_{\nu}$ by using Eq.~\eqref{eq:ISSneutrino}.

Also the inverse decays processes $\ell H\to N$, $\overline{\ell}H^* \to N$ can lead to a lepton number washout, if they feel lepton number violation. Indeed, if they do, then they distinguish the initial $\ell$ states from $\bar{\ell}$ ones and convert more $\ell$ than $\bar{\ell}$ back into $N$'s.
We can assume that the lepton asymmetry produced by $N$ decays does not get erased provided that the inverse-decays rate $\Gamma_{ID}\leq H$ at $T_\RH$. It was found by the authors of~\cite{Blanchet:2009kk} that the small mass splitting between the two states among the heavy pseudo-Dirac $N_i$ leads to a destructive interference between their contributions to $\Gamma_{ID}$, which has the effect of suppressing the washouts.
This can be parametrized by the suppression factor of the decay width
\begin{equation}
k^{\text{eff}}
=\frac{\rho^2}{1+\sqrt{c}\rho+\rho^2}
= \frac{M_N \mu^2}{\Gamma^2 M_N + \mu \Gamma^2 +M_N \mu^2}\, ,
\end{equation}
where $\rho \sim \mu/\Gamma$ and $c\sim (\Gamma/M_N)^2$.
The quadratic suppression in $\mu$ can be again understood from the fact that
inverse decays cannot wash out a lepton asymmetry in the lepton conserving limit, hence they should depend on some power of $\mu$ in addition to powers of the Yukawa couplings.
% the rate for $L$ violating processes should vanish in the lepton-number conserving limit.\\
One can thus estimate $\Gamma_{ID}$ as
\begin{equation}\label{eq:ISSonshell}
\Gamma_{ID}
\simeq 2 \Gamma\,k^{\text{eff}} \Big(\frac{M_N}{T_\RH}\Big)^{3/2}\text{Exp}\Big[-\frac{M_N}{T_\RH}\Big]\,,
\end{equation}
where the factor of 2 comes from the fact that both $N_1$ and $N_2$ contribute to washout.
This means that the rate of the net washout processes $\ell H\to N$, $\overline{\ell}H^* \to N$ at $T_\RH$ is suppressed by both the Boltzmann factor and the smallness of $\mu$. Therefore, washouts of the lepton asymmetry turns out to be very suppressed.
%and this makes our scenario potentially accessible at future experiments, as shown in Fig.~\ref{fig:parspaceISS}. 

To summarize we find that, in the model of leptogenesis from inverse-see-saw and composite neutrinos, $2\to2$ washout effects are never the leading constraint on the parameter space of our interest, while inverse decays can become an important constraint only if one lifts the Boltzmann suppression by choosing $M_N < 4\pi f$.

\subsection{Composite sterile neutrinos at colliders and discussion}
We are also interested in possible signatures of the sterile fermions $N$ at colliders, coming from their mixing with the active neutrinos.
The requirement $T_\RH > T_\text{Sph}$ plus $M_N \sim O(\pi f)$ implies $M_N \gtrsim$~TeV, making a future high-energy muon collider~\cite{AlAli:2021let,Aime:2022flm} the leading proposed collider to test this picture.
The authors of~\cite{Li:2023tbx} estimated the projected sensitivity of a $\sqrt{s}=10$ TeV muon collider with integrated luminosity 10 $\text{ab}^{-1}$
%on the squared mixing angle $|U_{\ell}|^2$, where $\ell$ refers to the lepton flavour, with the full-reconstructable HNL decay into a hadronic $W$ and a charged lepton.
on the squared mixing angle $|U_{\ell}|^2$ between the sterile $N$ and the SM neutrino of flavour $\ell$.
The dominant $N$-production mechanism is $\mu^+\mu^- \to N \bar{\nu}_\mu$, which, being $t$-channel, is not suppressed at large center-of-mass squared energies $s$, and the decay mode considered is $N \to W \ell$ which is fully reconstructable and whose branching ratio is approximately independent of $U_{\ell}$.

The best projected sensitivity occurs for the muon flavoured case and reaches $|U_{\mu}^2|\lesssim10^{-6}$ for a sterile fermion mass between 1 and 10 TeV, of course it does not extend to larger masses because $\sqrt{s}=10$ TeV. In the ISS model the active-sterile neutrino mixing is given by $\sin \theta \sim y\,v/(\sqrt{2}M_N)$, hence $|U_{\mu}^2|=\sin ^2\theta\sim y^2 v^2/(2 M_N^2)$.
Therefore, for fixed $y$, the sensitivity $|U_{\mu}^2|\lesssim10^{-6}$ translates into a lower limit on $M_N$: lighter values of $M_N$ will be potentially discoverable at a future high-energy muon collider, and our model turns out to be able to reproduce the BAU for those light $M_N$ values, compatibly with all other constraints.
%Therefore  we can fix a value of the Yukawa coupling $y$ compatible with the baryon yield and check whether there is some parameter space in the $\{\cvac, M_N\}$ plane in reach of the projected sensitivity, which is given by the dashed red vertical line in Fig.~\ref{fig:parspaceISS}. \\

\subsection{Summary of the results}

\begin{figure}[!hbt]
\centering
  \begin{minipage}[b]{0.48\textwidth}
   \includegraphics[height= 7cm, width=8cm]{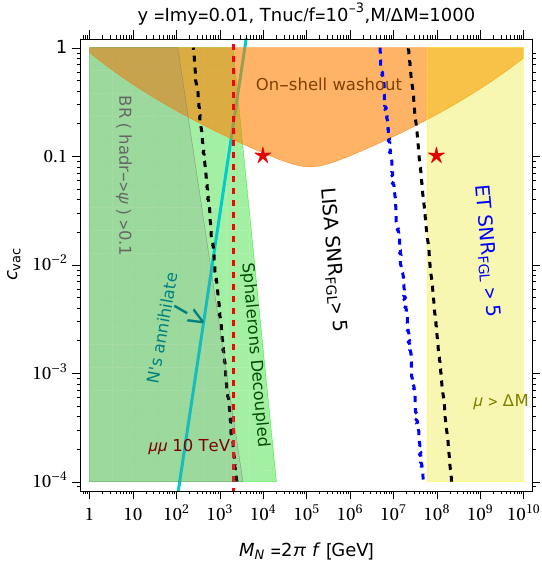}
   \end{minipage}
    \vspace{0.5cm}
    \hfill
 \begin{minipage}[b]{0.48\textwidth}
\includegraphics[height= 7cm, width=8cm]{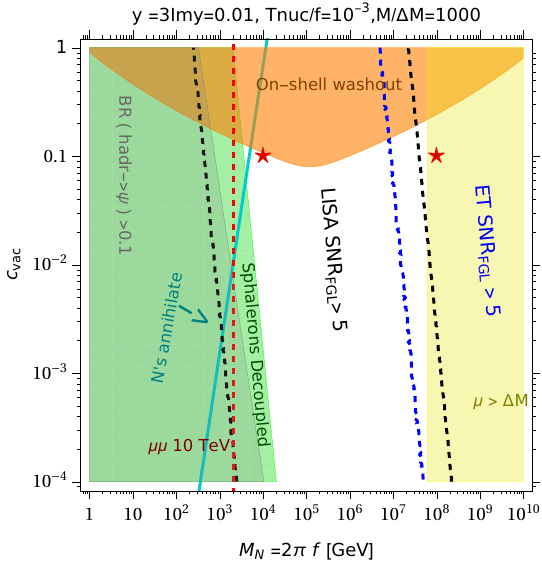}
  \end{minipage}
   \begin{minipage}[b]{0.48\textwidth}
   \includegraphics[height= 7cm, width=8cm]{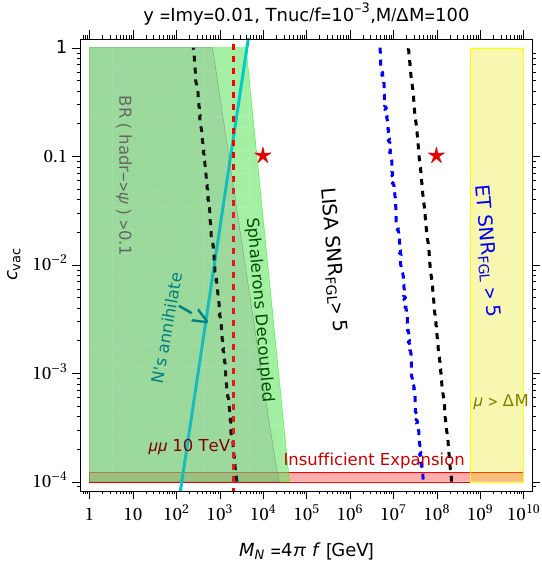}
    \end{minipage}
     \vspace{0.5cm}
    \hfill
 \begin{minipage}[b]{0.48\textwidth}
\includegraphics[height= 7cm, width=8cm]{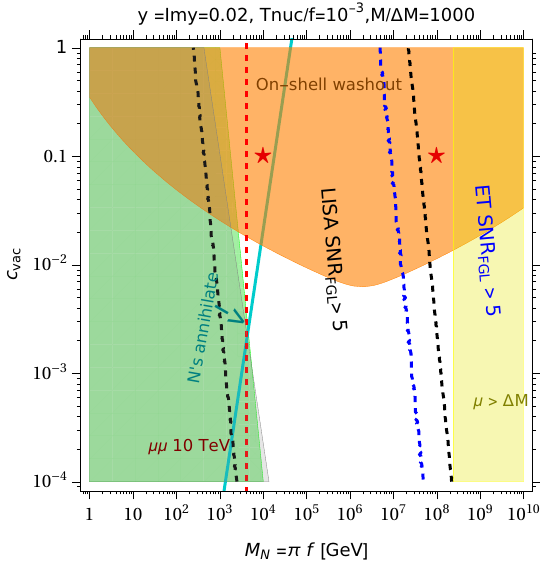}
  \end{minipage}
  \caption{
  Parameter space for the composite singlet fermion realisation of our framework, which constitutes a variation on inverse see-saw leptogenesis.  We choose $g_\TC = 80$, $g_* = g_\SM + g_\TC$, $g_\RH = g_\SM + 1$ and $Z=1$ for definiteness.
   The white area is allowed and accomodates the observed baryon asymmetry, the red and gray regions are excluded by the PT dynamics and above the cyan line the $N$'s start to equilibrate with the dilaton before they decay and our computation of the BAU should be changed, as in Fig.~\ref{fig:parspace_framework}.
  The dashed red line is obtained from the projected sensitivity of a 10 TeV muon collider on the active-sterile neutrino mixing, see~\cite{Li:2023tbx}. The green region is excluded from requiring electroweak sphalerons to be out of equilibrium for reheating temperature $T_\RH< 132$ GeV.
  In the orange exclusion regions the on-shell washout processes, i.e. inverse decays of Eq.~\eqref{eq:ISSonshell}, are in equilibrium at $T_\RH$.
 The yellow region is excluded by requiring $\mu<\Delta M$ for our expression of the rate asymmetry Eq.~\eqref{eq:ISSasymmetry} to hold. We also show the parameter space testable by the Einstein Telescope (ET) and LISA using  $\alpha_\GW=100$, $\beta/H=15$, where the astrophysical foregrounds have been taken into account. We denote with red stars the two points of the parameter space for which we computed the gravitational wave spectrum shown in Fig.~\ref{fig:gwspectrum}.}
    \label{fig:parspaceISS}
\end{figure}

We summarize our findings in Fig.~\ref{fig:parspaceISS}, where we set different benchmark values for $y, M_N/\Delta M$, $M_N/f$ and we take $\Tnuc/f=10^{-3}$ and $Z=1$. A remarkable feature of our mechanism is that, as it is manifest from the plots, one can accomplish leptogenesis around few TeV in the ISS scenario, while this is not the case for the standard thermal leptogenesis ISS scenario, where washout effects are active already at $T_\RH\sim M_N$ and are too strong to accomodate the required asymmetry, even taking into account quasi-degeneracy enhancements of the rate asymmetry, as shown in~\cite{Dolan:2018qpy,Agashe:2018cuf}. 

%One can see that we expect our model to be partially in reach of a 10 TeV muon collider.
%We find that some degeneracy between the $N_i$, of $\mathcal{O}(10^{2}-10^{3})$ and as expected given they are hadron from the same confining sector, is needed to reach the projected sensitivity of the muon collider.
One can see that we expect our model to be partially in reach of a 10 TeV muon collider, provided some mass degeneracy exists between the $N_i$, of $M_N/\Delta M \sim \mathcal{O}(10^{2}-10^{3})$ and as expected given they are hadrons from the same confining sector.
A detailed analysis would be needed to make this statement more precise, but goes beyond the scope of its work.
Coming to testability at lower-energy colliders, we find an obstacle to make our model viable at lower values of $f$.
The obstacle comes from the interplay between $N$ annihilations before decays and the bounds from washout effects and from the electroweak sphalerons. The latter need to be active, placing the absolute lower limit $T_\RH \gtrsim 132$~GeV, and hence demanding $f$ to be roughly larger than the weak scale.
One could then try to bring $M_N$ within reach of lower-energy colliders by choosing $M_N <(<) 4\pi f$, but washout effects would then kick-in to close the allowed parameter space, as visible in the Fig.~\ref{fig:parspaceISS}.
%\vdr{One can see that we expect our model to be only partially in reach of a 10 TeV muon collider, because of the interplay between the bounds from washout effects and from the electroweak sphalerons: while the former would be alleviated taking a large $M_N/f$ ratio, we would also prevent $M_N$ to be lighter than a few TeV, due to the lower bound on $T_\RH$ to have sphalerons in thermal equilibrium. On the other side if one tries to lower $M_N$ by reducing $M_N/f$, the washout from inverse decay becomes soon the dominant constraint due to Boltzmann suppression. Moreover we find that some degeneracy between the $N_i$ of $\mathcal{O}(10^{2}-10^{3})$ is needed to reach the projected sensitivity of the muon collider. However a detailed analysis would be needed to make this statement more precise.  }   
%

\subsection{A comment on linear see-saw}
What about other realizations of the seesaw mechanism, like the linear seesaw (LSS)?
In analogy to the ISS, additional sterile Dirac fermions $N_a$ are introduced, which mix with the SM neutrinos.
In the LSS $\mu = 0$ at tree level and lepton number is explicitly broken by an additional lepton number breaking Yukawa coupling $y'$ of $N_{aL}$ to the SM and we can assume without loss of generality that $y'<y$. The LSS Lagrangian is given by 
\begin{equation}
-\mathcal{L}^{LSS}= y_{a \alpha}\overline{N}_{aR} H L_{\alpha}+ M_{N_a}\overline{N}_a N_a +y'_{a \alpha}\overline{N}_{aL}^c H L_{\alpha}  +\hc ,
\end{equation}
and the light neutrinos mass matrix reads $m_{\nu}=v^2 (y'^T M_N^{T-1}y+ y^TM_N^{-1}y')$. A parametrical expression for the rate asymmetry $\epsilon^{LSS}$ was obtained in~\cite{Agashe:2018cuf}, $\epsilon^{LSS}\sim y'^2/(16\pi)$. Fixing now $y$ as a function of $y'$ and $ M_N$ by using the neutrino mass matrix relation and imposing for consistency that $y'<y$, one obtains an upper bound on $y'$,
\begin{equation}
y'< \Big(\frac{M_N m_{\nu}}{v^2}\Big)^{1/2}\sim 10^{-6} \Big(\frac{M_N}{\text{TeV}}\Big)^{1/2},
\end{equation}
which translates into a rate asymmetry $\epsilon^{LSS} \leq 10^{-12}$ around the TeV scale, way too small to account for the observed baryon yield. Therefore the pure LSS is not a viable model for TeV leptogenesis in our framework.

The discussion above however holds only at tree level. Starting from the pure LSS, a Majorana mass term $\mu$ for the $N_a$ is generated at the one-loop level, of the size $\mu\sim M_N yy'/(16\pi^2)$, thus one expects a general LSS+ISS model to arise from a LSS tree level Lagrangian~\cite{Agashe:2018cuf}.
We may then wonder if a new interesting region of the parameter space opens up in the region where the interference between ISS and LSS contributions produces the relevant term for the rate asymmetry: it turns out that the neutrino mass matrix formula is dominated by the LSS term,
while the rate asymmetry by the ISS piece and it reads
%and one can obtain the following parametrical expression for the rate asymmetry
%
\begin{equation}
\epsilon^{LSS+ISS}=\frac{m_{\nu}^2 M_N^2}{v^4 y^4}.
\label{eq:ISSLSSasymmetry}
\end{equation}
The $y$ coupling needed for leptogenesis, $y\sim \mathcal{O}(10^{-4}-10^{-5})$ for $M_N\sim$ TeV, is way too small to provide any signatures at colliders, so we will not explore in more detail this model within our supercooled confinement BAU framework.

\section{Conclusions and Outlook}
\label{sec:conclusions}
In this paper we have proposed a new framework for the generation of the observed baryon asymmetry of the universe (BAU), based on a first order supercooled phase transition (PT) of a confining sector.
In our framework $B-L$ violating processes are driven by the decays of some composite states (hadrons) of the confined phase.
%The main advantages with respect to the typical non-confining mechanism are that in our setup the rate asymmetry is enhanced  due to the deep inelastic scatterings occurring upon bubble percolation, which increase the population of the decaying hadrons, and that washout processes are greatly suppressed due to the typically large hierarchy between the hadron masses and the reheating temperature.  
The main advantages with respect to the analogous non-confining mechanism~\cite{Baldes:2021vyz} are that in our setup the rate asymmetry is enhanced  due to the deep inelastic scatterings occurring upon bubble percolation, which increase the population of the decaying hadrons, and due to the suppression of washout processes because of the hierarchy between the hadron masses and the reheating temperature.
Independently of its specific realization, our framework allows to reproduce the observed BAU down to the weak scale, and potentially down to the MeV one. A first important consequence of this result is that, contrarily to its weakly coupled realisations, the framework is testable in gravitational waves (GWs)  not only by the Einstein Telescope but also by lower frequency ones like BDECIGO and LISA. Based on the current understanding of astrophysical foregrounds, the absence of a signal in these experiments will exclude the framework down to the TeV scale, see Fig.~\ref{fig:parspace_framework}. 
%We pointed out that supercooled PTs are expected to produce a gravitational wave signal which is potentially in reach of current and future telescopes and then we specialized our framework to two specific models, already discussed in the literature, realizing the baryon asymmetry via a baryogenesis and leptogenesis mechanism respectively, in order to show that our framework extends their testability significantly. \\

We have then studied two specific models within our proposed framework, one realising baryogenesis from the decays of composite scalars, charged under the SM, one realising leptogenesis from the decays of composite fermions, singlets under the SM gauge. This allows to include effects that could potentially washout the BAU, to determine other experimental signals, and to possibly connect our framework with open SM problems in addition to the BAU.
\begin{itemize}
\item[$\diamond$]
Concerning baryogenesis from composite scalars, we have found that washout effects prevent PT scales $f$ lower than a few GeV, see Fig.~\ref{fig:parspace}. We have then studied collider searches for the composite scalars, finding they exclude values of $f$ lower than the weak scale, and that future LHC searches, like 4-jet ones, will test new open territory of this model. The model could also manifest itself in flavour and $CP$-violating observables, although the size of the effects depends on further details of the model and an absence of deviation from the SM would not exclude it.
Finally, we find that the dominant contribution to the neutron EDM arises at 3 loops and the model is not expected to show up there.
\item[$\diamond$]
Concerning leptogenesis from composite sterile neutrinos, it offers a natural implementation of inverse and linear see-saw scenarios (ISS and LSS), that also explains the small SM neutrino masses. 
Focusing on the ISS for definiteness, we found that we can reproduce the observed BAU down to scales of the PT $f \sim$ TeV, see Fig.~\ref{fig:parspaceISS}.
A degeneracy at the level of $10^{-2}$ or more is required between the masses of two composite neutrinos, but that is actually expected from a confining sector, while in the standard thermal leptogenesis one generally needs tuning.
The obstacle to reach $f$ below the electroweak scale comes from the requirement of a reheating temperature larger than about 130~GeV, so that sphalerons are active. Washout effects are further suppressed, with respect to a generic case, by the insertion of an extra parameter to break lepton number, and only inverse decays can lead to appreciable washout in case $M_N < 4\pi f$.
%We find that the model is partly testable by $N$ searches at a future muon collider.
We find that the model is barely testable by $N$ searches at a future muon collider.
Our results concerning this model are also interesting from the ISS and LSS points of view, because we enlarge the parameter space where these leptogenesis models work.

\end{itemize}
%
%For the former, realizing baryogenesis via the decay of a composite scalar with leptoquark and diquark couplings, we have shown that we can lower the mass of the decaying hadron down to the TeV scale, making the model potentially testable at colliders and at meson factories, with a gravitational wave signal in the reach of LISA and the Einstein Telescope. Other constraints from electric dipole moments and from washout suppressions are subdominant or can be easily evaded. \\

Our study opens several interesting avenues of further investigation, both on the model-building and on the phenomenology sides. Concerning the former, it would be for example interesting to connect our framework for the generation of the BAU with solutions to the hierarchy problem of the Fermi scale, given that it works down to the electroweak scale. Concerning phenomenology, it would be interesting to find a specific model realising our framework which allows to go down to the MeV scale for the PT, so to possibly explain the GWs observed by pulsar timing arrays~\cite{NANOGrav:2020bcs,Goncharov:2021oub,EPTA:2021crs,Antoniadis:2022pcn}. We leave these and other directions for future work.

\section*{Acknowledgements}
We thank Iason Baldes for useful discussions and comments on the manuscript, and Yann Gouttenoire for useful discussions.
This work was supported in part by the European Union’s Horizon research and innovation programme under the Marie Sklodowska-Curie grant agreements No. 860881-HIDDeN
and No. 101086085-ASYMMETRY, by the Italian INFN program on Theoretical Astroparticle Physics, and by the French CNRS grant IEA “DaCo: Dark Connections”.

\appendix

\section{Cancellation of the 2-loop contributions to the nEDM} \label{sec:app1}

 \begin{figure}[!hbt] 
\raggedleft
  \includegraphics%[height=4.5cm, width=18cm]
  [width=\textwidth]{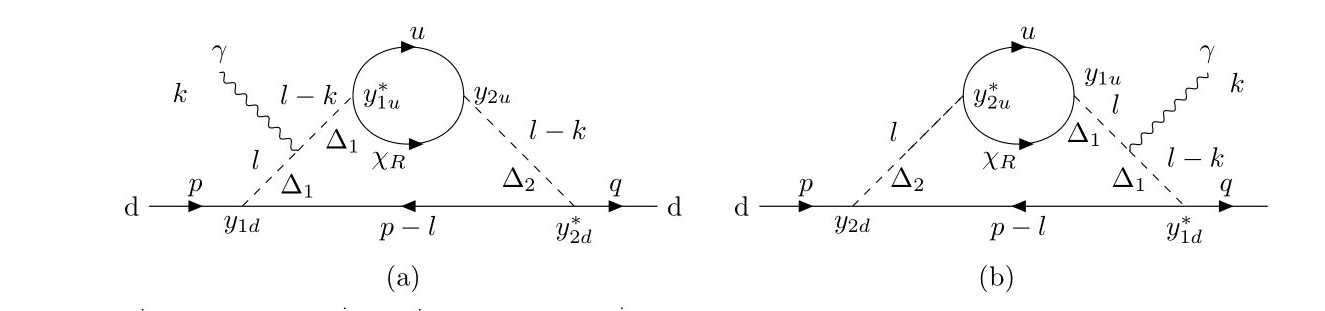}
\caption{Two-loop conjugate diagrams contributing to the nEDM. In order to obtain a cancellation one needs to take into account also the pair of diagrams where $\Delta_1$ and $\Delta_2$ are interchanged. }\label{fig:twoloop}\end{figure}

Let us show explicitly that the sum of the diagrams shown in Fig.~\ref{fig:twoloop} with in addition those with $\Delta_{1,2}$ swapped provides a vanishing contribution to the nEDM. \\
The starting point is given by the structure of the electromagnetic nucleon vertex in presence of CP-violation, which is given by 
\begin{equation}
i\mathcal{M}^{\mu}=\overline {u_N}(q)\Big[\gamma^{\mu} F_E(k^2)+\frac{i\sigma^{\mu\nu}k_{\nu}}{2 m_N} F_M(k^2)+ \frac{i\sigma^{\mu\nu}k_{\nu}}{2 m_N} \gamma_5 F_D(k^2) \Big]u_N(p).
\end{equation}
Here $k=(p-q)$ is the momentum of the photon, $m_N$ is the nucleon mass and the EDM of the nucleon is defined as $d_n=F_D(0)/(2m_N)$. Using the Gordon-decomposition of the vector-axial current, the CP violating ($\cancel{CP}$) part of the above expression reads
\begin{equation}
i \mathcal{M}^{\mu}_{\cancel{CP}}=d_n \overline{u_N}(q) \gamma_5  u_N(p) (p+q)^{\mu}
\end{equation} 
Thus we need to concentrate on the $(p+q)\cdot \epsilon$ terms in the computation of the EDM, where $\epsilon$ is the photon polarization. \\
In order to find whether there is a cancellation among the diagrams, we can factor out the upper loop, as it is symmetric under $\Delta_{1,2}$ exchange. Let's  focus thus on the lower loop and denote by $Y$ the product of all the Yukawa couplings entering the diagram (a). The relevant amplitude reads
\begin{equation}\begin{split}
iM=Y \int \frac{d^{4}\ell}{(2\pi)^{4}}\overline{u}(q) P_L \frac{i}{\slashed{p}-\slashed{\ell}-m}P_R u(p)\frac{i}{\ell^2-M_1^{2}}\frac{i}{(\ell-k)^2-M_1^{2}}\Big(\frac{i(2 \ell-k)\cdot \epsilon (k)}{(\ell-k)^2-M_2^2}\Big)  \\
=Y \overline{u}(q) \gamma^\alpha P_R u(p)\int \frac{d^{4}\ell}{(2\pi)^{4}}\frac{(p^\alpha -\ell^\alpha) \ell \cdot \epsilon (k) }{(\ell^2-M_1^{2})((\ell-k)^2-M_1^{2})((\ell-k)^2-M_2^2)((p-\ell)^2-m^2)}
\end{split}\end{equation}
where $m$ is the mass of the internal quark and $M_{1,2}$ the masses of $\Delta_{1,2}$. Since the $\Delta_i$ live at much higher scales than those probed by the EDM we can assume $M_{1,2}^2\gg k^2, p^2,m^2$. The integrand is dominated by $\ell^2 \sim M_{1,2}^2$, thus one can expand the propagators  as
\begin{equation}
\frac{1}{(\ell-k)^2-M_1^2}\simeq \frac{1}{\ell^2-M_1^2}\Big[1+ 2\frac{ \ell\cdot k}{\ell^2-M_1^2}+4 \frac{(\ell \cdot k)^2}{(\ell^2-M_1^2)^2}\Big]\, ,
\end{equation}
with analogous expressions for the other propagators. One obtains
\begin{equation}\begin{split}
iM= Y \overline{u}(q) \gamma^\alpha P_R u(p)\int \frac{d^{4}\ell}{(2\pi)^{4}}\frac{(p^\alpha -\ell^\alpha) \ell \cdot \epsilon \Big[1+ 2\frac{ \ell\cdot k}{\ell^2-M_1^2}+4 \frac{(\ell \cdot k)^2}{(\ell^2-M_1^2)^2}\Big]\Big[1+ 2\frac{ \ell\cdot k}{\ell^2-M_2^2}+4 \frac{(\ell \cdot k)^2}{(\ell^2-M_2^2)^2}\Big]}{\ell^2 (l^2-M_1^{2})^2(\ell^2-M_2^2)}\\ \Big[1+ 2\frac{ \ell\cdot p}{\ell^2}+4 \frac{(\ell \cdot p)^2}{\ell^4}\Big]
\end{split} \end{equation}
The denominator is now an even function of the loop momentum $\ell$, thus we need to focus on terms with an even number of $\ell$ at numerator. At $\mathcal {O} (\ell^2)$ there are two kind of terms which might contribute to the nEDM,
\begin{itemize}
\item $p^{\alpha}\,\ell^\mu\,\ell^\beta \,p_\beta \, \epsilon_\mu \sim \ell^2 \,p^\alpha \,(p \cdot \epsilon)$
\item  $p^{\alpha}\,\ell^\mu\,\ell^\beta k_\beta  \, \epsilon_\mu \sim  \ell^2 \,p^\alpha \,(k \cdot \epsilon)=0$
\end{itemize}
and as the non-zero one appears also in diagram (b) the sum of the contribution of (a) and (b) to the nEDM would cancel. Let us consider now $\mathcal{O}(\ell^4)$ terms.  There are three types of terms which might contribute: 
\begin{itemize}
\item  $\ell ^\alpha \,\ell ^\mu \,\ell ^\beta \,\ell ^\gamma \,  k_\beta  \, k_\gamma \, \epsilon_\mu \sim (\ell^2)^2 \Big[\epsilon_\alpha\,k^2+ 2\, k_\alpha (k\cdot \epsilon)\Big]=0$
\item    $\ell ^\alpha \,\ell ^\mu \,\ell ^\beta \,\ell ^\gamma \,  p_\beta  \, p_\gamma \, \epsilon_\mu \sim (\ell^2)^2 \Big[\epsilon_\alpha\,p^2+ 2\, p_\alpha (p\cdot \epsilon)\Big]$
\item   $\ell ^\alpha \,\ell ^\mu \,\ell ^\beta \,\ell ^\gamma \,  k_\beta    \, p_\gamma \, \epsilon_\mu \sim (\ell^2)^2 \Big[\epsilon_\alpha (k \cdot p)+  p_\alpha (k\cdot \epsilon)+ k_\alpha (p\cdot \epsilon)\Big]\sim (\ell^2)^2 \, k_\alpha (p\cdot \epsilon)$
\end{itemize}
The second term appears also in the diagram (b) with coefficient $Y^*$ and so it is harmless. The third one receives two contributions from diagram (a), from $\Delta_1$ and $\Delta_2$,  while (b) contains only the $\Delta_1$ contribution, as the momentum flowing in $\Delta_2$ doesn't depend on $k$. Therefore the sum of diagrams (a) and (b) doesn't provide a non-zero nEDM. However, the surviving contribution stems from an integral with a $1/(\ell^2-M_1^2)^2(\ell^2-M_2^2)^2$ factor and is cancelled by the couple of conjugate diagrams with $\Delta_{1,2}$ exchanged. We can thus conclude that our model allows new physics contributions to the nEDM to occur only starting at the three loop level.

\bibliographystyle{utphys}
\bibliography{refs}

\providecommand{\href}[2]{#2}\begingroup\raggedright\begin{thebibliography}{100}

\bibitem{Planck:2018vyg}
{\bfseries Planck} Collaboration, N.~Aghanim {\em et~al.}, ``{Planck 2018
  results. VI. Cosmological parameters},''
  \href{http://dx.doi.org/10.1051/0004-6361/201833910}{{\em Astron. Astrophys.}
  {\bfseries 641} (2020) A6}, \href{http://arxiv.org/abs/1807.06209}{{\ttfamily
  arXiv:1807.06209 [astro-ph.CO]}}. [Erratum: Astron.Astrophys. 652, C4
  (2021)].

\bibitem{Fields:2019pfx}
B.~D. Fields, K.~A. Olive, T.-H. Yeh, and C.~Young, ``{Big-Bang Nucleosynthesis
  after Planck},'' \href{http://dx.doi.org/10.1088/1475-7516/2020/03/010}{{\em
  JCAP} {\bfseries 03} (2020) 010},
  \href{http://arxiv.org/abs/1912.01132}{{\ttfamily arXiv:1912.01132
  [astro-ph.CO]}}. [Erratum: JCAP 11, E02 (2020)].

\bibitem{Cohen:1997ac}
A.~G. Cohen, A.~De~Rujula, and S.~L. Glashow, ``{A Matter - antimatter
  universe?},'' \href{http://dx.doi.org/10.1086/305328}{{\em Astrophys. J.}
  {\bfseries 495} (1998) 539--549},
  \href{http://arxiv.org/abs/astro-ph/9707087}{{\ttfamily
  arXiv:astro-ph/9707087}}.

\bibitem{Steigman:2008ap}
G.~Steigman, ``{When Clusters Collide: Constraints On Antimatter On The Largest
  Scales},'' \href{http://dx.doi.org/10.1088/1475-7516/2008/10/001}{{\em JCAP}
  {\bfseries 10} (2008) 001}, \href{http://arxiv.org/abs/0808.1122}{{\ttfamily
  arXiv:0808.1122 [astro-ph]}}.

\bibitem{Kolb:1990vq}
E.~W. Kolb and M.~S. Turner,
  \href{http://dx.doi.org/10.1201/9780429492860}{{\em {The Early Universe}}},
  vol.~69.
\newblock 1990.

\bibitem{Sakharov:1967dj}
A.~D. Sakharov, ``{Violation of CP Invariance, C asymmetry, and baryon
  asymmetry of the universe},''
  \href{http://dx.doi.org/10.1070/PU1991v034n05ABEH002497}{{\em Pisma Zh. Eksp.
  Teor. Fiz.} {\bfseries 5} (1967) 32--35}.

\bibitem{Aoki:2006we}
Y.~Aoki, G.~Endrodi, Z.~Fodor, S.~D. Katz, and K.~K. Szabo, ``{The Order of the
  quantum chromodynamics transition predicted by the standard model of particle
  physics},'' \href{http://dx.doi.org/10.1038/nature05120}{{\em Nature}
  {\bfseries 443} (2006) 675--678},
  \href{http://arxiv.org/abs/hep-lat/0611014}{{\ttfamily
  arXiv:hep-lat/0611014}}.

\bibitem{Kajantie:1996mn}
K.~Kajantie, M.~Laine, K.~Rummukainen, and M.~E. Shaposhnikov, ``{Is there a~
  hot electroweak phase transition at $m_H \gtrsim m_W$?},''
  \href{http://dx.doi.org/10.1103/PhysRevLett.77.2887}{{\em Phys. Rev. Lett.}
  {\bfseries 77} (1996) 2887--2890},
  \href{http://arxiv.org/abs/hep-ph/9605288}{{\ttfamily arXiv:hep-ph/9605288}}.

\bibitem{Creminelli:2001th}
P.~Creminelli, A.~Nicolis, and R.~Rattazzi, ``{Holography and the electroweak
  phase transition},''
  \href{http://dx.doi.org/10.1088/1126-6708/2002/03/051}{{\em JHEP} {\bfseries
  03} (2002) 051},
\href{http://arxiv.org/abs/hep-th/0107141}{{\ttfamily arXiv:hep-th/0107141
  [hep-th]}}.
%%CITATION = HEP-TH/0107141;%%.

\bibitem{Nardini:2007me}
G.~Nardini, M.~Quiros, and A.~Wulzer, ``{A Confining Strong First-Order
  Electroweak Phase Transition},''
  \href{http://dx.doi.org/10.1088/1126-6708/2007/09/077}{{\em JHEP} {\bfseries
  09} (2007) 077},
\href{http://arxiv.org/abs/0706.3388}{{\ttfamily arXiv:0706.3388 [hep-ph]}}.
%%CITATION = ARXIV:0706.3388;%%.

\bibitem{Konstandin:2011dr}
T.~Konstandin and G.~Servant, ``{Cosmological Consequences of Nearly Conformal
  Dynamics at the TeV scale},''
  \href{http://dx.doi.org/10.1088/1475-7516/2011/12/009}{{\em JCAP} {\bfseries
  1112} (2011) 009},
\href{http://arxiv.org/abs/1104.4791}{{\ttfamily arXiv:1104.4791 [hep-ph]}}.
%%CITATION = ARXIV:1104.4791;%%.

\bibitem{Craig:2020jfv}
N.~Craig, N.~Levi, A.~Mariotti, and D.~Redigolo, ``{Ripples in Spacetime from
  Broken Supersymmetry},''
  \href{http://dx.doi.org/10.1007/JHEP02(2021)184}{{\em JHEP} {\bfseries 21}
  (2020) 184}, \href{http://arxiv.org/abs/2011.13949}{{\ttfamily
  arXiv:2011.13949 [hep-ph]}}.

\bibitem{Jinno:2016knw}
R.~Jinno and M.~Takimoto, ``{Probing a classically conformal B-L model with
  gravitational waves},''
  \href{http://dx.doi.org/10.1103/PhysRevD.95.015020}{{\em Phys. Rev. D}
  {\bfseries 95} no.~1, (2017) 015020},
  \href{http://arxiv.org/abs/1604.05035}{{\ttfamily arXiv:1604.05035
  [hep-ph]}}.

\bibitem{DelleRose:2019pgi}
L.~Delle~Rose, G.~Panico, M.~Redi, and A.~Tesi, ``{Gravitational Waves from
  Supercool Axions},'' \href{http://dx.doi.org/10.1007/JHEP04(2020)025}{{\em
  JHEP} {\bfseries 04} (2020) 025},
  \href{http://arxiv.org/abs/1912.06139}{{\ttfamily arXiv:1912.06139
  [hep-ph]}}.

\bibitem{VonHarling:2019rgb}
B.~Von~Harling, A.~Pomarol, O.~Pujol\`as, and F.~Rompineve, ``{Peccei-Quinn
  Phase Transition at LIGO},''
  \href{http://dx.doi.org/10.1007/JHEP04(2020)195}{{\em JHEP} {\bfseries 04}
  (2020) 195}, \href{http://arxiv.org/abs/1912.07587}{{\ttfamily
  arXiv:1912.07587 [hep-ph]}}.

\bibitem{Greljo:2019xan}
A.~Greljo, T.~Opferkuch, and B.~A. Stefanek, ``{Gravitational Imprints of
  Flavor Hierarchies},''
  \href{http://dx.doi.org/10.1103/PhysRevLett.124.171802}{{\em Phys. Rev.
  Lett.} {\bfseries 124} no.~17, (2020) 171802},
  \href{http://arxiv.org/abs/1910.02014}{{\ttfamily arXiv:1910.02014
  [hep-ph]}}.

\bibitem{Bodeker:2020ghk}
D.~Bodeker and W.~Buchmuller, ``{Baryogenesis from the weak scale to the grand
  unification scale},''
  \href{http://dx.doi.org/10.1103/RevModPhys.93.035004}{{\em Rev. Mod. Phys.}
  {\bfseries 93} no.~3, (2021) 035004},
  \href{http://arxiv.org/abs/2009.07294}{{\ttfamily arXiv:2009.07294
  [hep-ph]}}.

\bibitem{Witten:1984rs}
E.~Witten, ``{Cosmic Separation of Phases},''
\href{http://dx.doi.org/10.1103/PhysRevD.30.272}{{\em Phys. Rev.} {\bfseries
  D30} (1984) 272--285}.
%%CITATION = PHRVA,D30,272;%%.

\bibitem{Hogan:1986qda}
C.~J. Hogan, ``{Gravitational radiation from cosmological phase transitions},''
{\em Mon. Not. Roy. Astron. Soc.} {\bfseries 218} (1986) 629--636.
%%CITATION = MNRAA,218,629;%%.

\bibitem{NANOGrav:2023hvm}
{\bfseries NANOGrav} Collaboration, A.~Afzal {\em et~al.}, ``{The NANOGrav 15
  yr Data Set: Search for Signals from New Physics},''
  \href{http://dx.doi.org/10.3847/2041-8213/acdc91}{{\em Astrophys. J. Lett.}
  {\bfseries 951} no.~1, (2023) L11},
  \href{http://arxiv.org/abs/2306.16219}{{\ttfamily arXiv:2306.16219
  [astro-ph.HE]}}.

\bibitem{Antoniadis:2023zhi}
J.~Antoniadis {\em et~al.}, ``{The second data release from the European Pulsar
  Timing Array: V. Implications for massive black holes, dark matter and the
  early Universe},'' \href{http://arxiv.org/abs/2306.16227}{{\ttfamily
  arXiv:2306.16227 [astro-ph.CO]}}.

\bibitem{Gouttenoire:2023bqy}
Y.~Gouttenoire, ``{First-Order Phase Transition Interpretation of Pulsar Timing
  Array Signal Is Consistent with Solar-Mass Black Holes},''
  \href{http://dx.doi.org/10.1103/PhysRevLett.131.171404}{{\em Phys. Rev.
  Lett.} {\bfseries 131} no.~17, (2023) 171404},
  \href{http://arxiv.org/abs/2307.04239}{{\ttfamily arXiv:2307.04239
  [hep-ph]}}.

\bibitem{LISACosmologyWorkingGroup:2022jok}
{\bfseries LISA Cosmology Working Group} Collaboration, P.~Auclair {\em
  et~al.}, ``{Cosmology with the Laser Interferometer Space Antenna},''
  \href{http://arxiv.org/abs/2204.05434}{{\ttfamily arXiv:2204.05434
  [astro-ph.CO]}}.

\bibitem{Falkowski:2012fb}
A.~Falkowski and J.~M. No, ``{Non-thermal Dark Matter Production from the
  Electroweak Phase Transition: Multi-TeV WIMPs and 'Baby-Zillas'},''
  \href{http://dx.doi.org/10.1007/JHEP02(2013)034}{{\em JHEP} {\bfseries 02}
  (2013) 034},
\href{http://arxiv.org/abs/1211.5615}{{\ttfamily arXiv:1211.5615 [hep-ph]}}.
%%CITATION = ARXIV:1211.5615;%%.

\bibitem{Hambye:2018qjv}
T.~Hambye, A.~Strumia, and D.~Teresi, ``{Super-cool Dark Matter},''
  \href{http://dx.doi.org/10.1007/JHEP08(2018)188}{{\em JHEP} {\bfseries 08}
  (2018) 188},
\href{http://arxiv.org/abs/1805.01473}{{\ttfamily arXiv:1805.01473 [hep-ph]}}.
%%CITATION = ARXIV:1805.01473;%%.

\bibitem{Baldes:2020kam}
I.~Baldes, Y.~Gouttenoire, and F.~Sala, ``{String Fragmentation in Supercooled
  Confinement and Implications for Dark Matter},''
  \href{http://dx.doi.org/10.1007/JHEP04(2021)278}{{\em JHEP} {\bfseries 04}
  (2021) 278}, \href{http://arxiv.org/abs/2007.08440}{{\ttfamily
  arXiv:2007.08440 [hep-ph]}}.

\bibitem{Azatov:2021ifm}
A.~Azatov, M.~Vanvlasselaer, and W.~Yin, ``{Dark Matter production from
  relativistic bubble walls},''
  \href{http://dx.doi.org/10.1007/JHEP03(2021)288}{{\em JHEP} {\bfseries 03}
  (2021) 288}, \href{http://arxiv.org/abs/2101.05721}{{\ttfamily
  arXiv:2101.05721 [hep-ph]}}.

\bibitem{Baldes:2022oev}
I.~Baldes, Y.~Gouttenoire, and F.~Sala, ``{Hot and heavy dark matter from a
  weak scale phase transition},''
  \href{http://dx.doi.org/10.21468/SciPostPhys.14.3.033}{{\em SciPost Phys.}
  {\bfseries 14} (2023) 033}, \href{http://arxiv.org/abs/2207.05096}{{\ttfamily
  arXiv:2207.05096 [hep-ph]}}.

\bibitem{Baldes:2023fsp}
I.~Baldes, M.~Dichtl, Y.~Gouttenoire, and F.~Sala, ``{Bubbletrons},''
  \href{http://arxiv.org/abs/2306.15555}{{\ttfamily arXiv:2306.15555
  [hep-ph]}}.

\bibitem{Liu:2021svg}
J.~Liu, L.~Bian, R.-G. Cai, Z.-K. Guo, and S.-J. Wang, ``{Primordial Black Hole
  Production during First-Order Phase Transitions},''
  \href{http://dx.doi.org/10.1103/PhysRevD.105.L021303}{{\em Phys. Rev. D}
  {\bfseries 105} no.~2, (2022) L021303},
  \href{http://arxiv.org/abs/2106.05637}{{\ttfamily arXiv:2106.05637
  [astro-ph.CO]}}.

\bibitem{Hashino:2021qoq}
K.~Hashino, S.~Kanemura, and T.~Takahashi, ``{Primordial black holes as a probe
  of strongly first-order electroweak phase transition},''
  \href{http://dx.doi.org/10.1016/j.physletb.2022.137261}{{\em Phys. Lett. B}
  {\bfseries 833} (2022) 137261},
  \href{http://arxiv.org/abs/2111.13099}{{\ttfamily arXiv:2111.13099
  [hep-ph]}}.

\bibitem{Lewicki:2023ioy}
M.~Lewicki, P.~Toczek, and V.~Vaskonen, ``{Primordial black holes from strong
  first-order phase transitions},''
  \href{http://dx.doi.org/10.1007/JHEP09(2023)092}{{\em JHEP} {\bfseries 09}
  (2023) 092}, \href{http://arxiv.org/abs/2305.04924}{{\ttfamily
  arXiv:2305.04924 [astro-ph.CO]}}.

\bibitem{Gouttenoire:2023naa}
Y.~Gouttenoire and T.~Volansky, ``{Primordial Black Holes from Supercooled
  Phase Transitions},'' \href{http://arxiv.org/abs/2305.04942}{{\ttfamily
  arXiv:2305.04942 [hep-ph]}}.

\bibitem{Katz:2016adq}
A.~Katz and A.~Riotto, ``{Baryogenesis and Gravitational Waves from Runaway
  Bubble Collisions},''
  \href{http://dx.doi.org/10.1088/1475-7516/2016/11/011}{{\em JCAP} {\bfseries
  1611} no.~11, (2016) 011},
\href{http://arxiv.org/abs/1608.00583}{{\ttfamily arXiv:1608.00583 [hep-ph]}}.
%%CITATION = ARXIV:1608.00583;%%.

\bibitem{Azatov:2021irb}
A.~Azatov, M.~Vanvlasselaer, and W.~Yin, ``{Baryogenesis via Relativistic
  Bubble Walls},'' \href{http://dx.doi.org/10.1007/JHEP10(2021)043}{{\em JHEP}
  {\bfseries 10} (2021) 043}, \href{http://arxiv.org/abs/2106.14913}{{\ttfamily
  arXiv:2106.14913 [hep-ph]}}.

\bibitem{Baldes:2021vyz}
I.~Baldes, S.~Blasi, A.~Mariotti, A.~Sevrin, and K.~Turbang, ``{Baryogenesis
  via relativistic bubble expansion},''
  \href{http://dx.doi.org/10.1103/PhysRevD.104.115029}{{\em Phys. Rev. D}
  {\bfseries 104} no.~11, (2021) 115029},
  \href{http://arxiv.org/abs/2106.15602}{{\ttfamily arXiv:2106.15602
  [hep-ph]}}.

\bibitem{Huang:2022vkf}
P.~Huang and K.-P. Xie, ``{Leptogenesis triggered by a first-order phase
  transition},'' \href{http://dx.doi.org/10.1007/JHEP09(2022)052}{{\em JHEP}
  {\bfseries 09} (2022) 052}, \href{http://arxiv.org/abs/2206.04691}{{\ttfamily
  arXiv:2206.04691 [hep-ph]}}.

\bibitem{Chun:2023ezg}
E.~J. Chun, T.~P. Dutka, T.~H. Jung, X.~Nagels, and M.~Vanvlasselaer,
  ``{Bubble-assisted leptogenesis},''
  \href{http://dx.doi.org/10.1007/JHEP09(2023)164}{{\em JHEP} {\bfseries 09}
  (2023) 164}, \href{http://arxiv.org/abs/2305.10759}{{\ttfamily
  arXiv:2305.10759 [hep-ph]}}.

\bibitem{Hindmarsh:2020hop}
M.~B. Hindmarsh, M.~L\"uben, J.~Lumma, and M.~Pauly, ``{Phase transitions in
  the early universe},''
  \href{http://dx.doi.org/10.21468/SciPostPhysLectNotes.24}{{\em SciPost Phys.
  Lect. Notes} {\bfseries 24} (2021) 1},
  \href{http://arxiv.org/abs/2008.09136}{{\ttfamily arXiv:2008.09136
  [astro-ph.CO]}}.

\bibitem{Gouttenoire:2022gwi}
Y.~Gouttenoire, ``{Beyond the Standard Model Cocktail},''
  \href{http://arxiv.org/abs/2207.01633}{{\ttfamily arXiv:2207.01633
  [hep-ph]}}.

\bibitem{Witten:1980ez}
E.~Witten, ``{Cosmological Consequences of a Light Higgs Boson},''
  \href{http://dx.doi.org/10.1016/0550-3213(81)90182-6}{{\em Nucl. Phys. B}
  {\bfseries 177} (1981) 477--488}.

\bibitem{Levi:2022bzt}
N.~Levi, T.~Opferkuch, and D.~Redigolo, ``{The supercooling window at weak and
  strong coupling},'' \href{http://dx.doi.org/10.1007/JHEP02(2023)125}{{\em
  JHEP} {\bfseries 02} (2023) 125},
  \href{http://arxiv.org/abs/2212.08085}{{\ttfamily arXiv:2212.08085
  [hep-ph]}}.

\bibitem{Coleman:1973jx}
S.~R. Coleman and E.~J. Weinberg, ``{Radiative Corrections as the Origin of
  Spontaneous Symmetry Breaking},''
  \href{http://dx.doi.org/10.1103/PhysRevD.7.1888}{{\em Phys. Rev. D}
  {\bfseries 7} (1973) 1888--1910}.

\bibitem{Gildener:1976ih}
E.~Gildener and S.~Weinberg, ``{Symmetry Breaking and Scalar Bosons},''
  \href{http://dx.doi.org/10.1103/PhysRevD.13.3333}{{\em Phys. Rev. D}
  {\bfseries 13} (1976) 3333}.

\bibitem{Randall:2006py}
L.~Randall and G.~Servant, ``{Gravitational waves from warped spacetime},''
  \href{http://dx.doi.org/10.1088/1126-6708/2007/05/054}{{\em JHEP} {\bfseries
  05} (2007) 054}, \href{http://arxiv.org/abs/hep-ph/0607158}{{\ttfamily
  arXiv:hep-ph/0607158}}.

\bibitem{Baldes:2021aph}
I.~Baldes, Y.~Gouttenoire, F.~Sala, and G.~Servant, ``{Supercool composite Dark
  Matter beyond 100 TeV},''
  \href{http://dx.doi.org/10.1007/JHEP07(2022)084}{{\em JHEP} {\bfseries 07}
  (2022) 084}, \href{http://arxiv.org/abs/2110.13926}{{\ttfamily
  arXiv:2110.13926 [hep-ph]}}.

\bibitem{Iso:2017uuu}
S.~Iso, P.~D. Serpico, and K.~Shimada, ``{QCD-Electroweak First-Order Phase
  Transition in a Supercooled Universe},''
  \href{http://dx.doi.org/10.1103/PhysRevLett.119.141301}{{\em Phys. Rev.
  Lett.} {\bfseries 119} no.~14, (2017) 141301},
  \href{http://arxiv.org/abs/1704.04955}{{\ttfamily arXiv:1704.04955
  [hep-ph]}}.

\bibitem{vonHarling:2017yew}
B.~von Harling and G.~Servant, ``{QCD-induced Electroweak Phase Transition},''
  \href{http://dx.doi.org/10.1007/JHEP01(2018)159}{{\em JHEP} {\bfseries 01}
  (2018) 159}, \href{http://arxiv.org/abs/1711.11554}{{\ttfamily
  arXiv:1711.11554 [hep-ph]}}.

\bibitem{Contino:2010unpub}
R.~Contino, A.~Pomarol, and R.~Rattazzi, ``{Unpublished work},''.
\newblock
  \url{https://indico.cern.ch/event/75810/contributions/1250635/attachments/1050757/1498158/Rattazzi.pdf}.

\bibitem{Appelquist:2010gy}
T.~Appelquist and Y.~Bai, ``{A Light Dilaton in Walking Gauge Theories},''
  \href{http://dx.doi.org/10.1103/PhysRevD.82.071701}{{\em Phys. Rev. D}
  {\bfseries 82} (2010) 071701},
  \href{http://arxiv.org/abs/1006.4375}{{\ttfamily arXiv:1006.4375 [hep-ph]}}.

\bibitem{Bellazzini:2012vz}
B.~Bellazzini, C.~Csaki, J.~Hubisz, J.~Serra, and J.~Terning, ``{A Higgslike
  Dilaton},'' \href{http://dx.doi.org/10.1140/epjc/s10052-013-2333-x}{{\em Eur.
  Phys. J. C} {\bfseries 73} no.~2, (2013) 2333},
  \href{http://arxiv.org/abs/1209.3299}{{\ttfamily arXiv:1209.3299 [hep-ph]}}.

\bibitem{Coradeschi:2013gda}
F.~Coradeschi, P.~Lodone, D.~Pappadopulo, R.~Rattazzi, and L.~Vitale, ``{A
  naturally light dilaton},''
  \href{http://dx.doi.org/10.1007/JHEP11(2013)057}{{\em JHEP} {\bfseries 11}
  (2013) 057}, \href{http://arxiv.org/abs/1306.4601}{{\ttfamily arXiv:1306.4601
  [hep-th]}}.

\bibitem{Chacko:2013dra}
Z.~Chacko, R.~K. Mishra, and D.~Stolarski, ``{Dynamics of a Stabilized Radion
  and Duality},'' \href{http://dx.doi.org/10.1007/JHEP09(2013)121}{{\em JHEP}
  {\bfseries 09} (2013) 121}, \href{http://arxiv.org/abs/1304.1795}{{\ttfamily
  arXiv:1304.1795 [hep-ph]}}.

\bibitem{Megias:2014iwa}
E.~Megias and O.~Pujolas, ``{Naturally light dilatons from nearly marginal
  deformations},'' \href{http://dx.doi.org/10.1007/JHEP08(2014)081}{{\em JHEP}
  {\bfseries 08} (2014) 081}, \href{http://arxiv.org/abs/1401.4998}{{\ttfamily
  arXiv:1401.4998 [hep-th]}}.

\bibitem{Konstandin:2010dm}
T.~Konstandin and J.~M. No, ``{Hydrodynamic obstruction to bubble expansion},''
  \href{http://dx.doi.org/10.1088/1475-7516/2011/02/008}{{\em JCAP} {\bfseries
  02} (2011) 008}, \href{http://arxiv.org/abs/1011.3735}{{\ttfamily
  arXiv:1011.3735 [hep-ph]}}.

\bibitem{Laurent:2022jrs}
B.~Laurent and J.~M. Cline, ``{First principles determination of bubble wall
  velocity},'' \href{http://dx.doi.org/10.1103/PhysRevD.106.023501}{{\em Phys.
  Rev. D} {\bfseries 106} no.~2, (2022) 023501},
  \href{http://arxiv.org/abs/2204.13120}{{\ttfamily arXiv:2204.13120
  [hep-ph]}}.

\bibitem{Bodeker:2009qy}
D.~Bodeker and G.~D. Moore, ``{Can electroweak bubble walls run away?},''
  \href{http://dx.doi.org/10.1088/1475-7516/2009/05/009}{{\em JCAP} {\bfseries
  05} (2009) 009}, \href{http://arxiv.org/abs/0903.4099}{{\ttfamily
  arXiv:0903.4099 [hep-ph]}}.

\bibitem{Bodeker:2017cim}
D.~Bodeker and G.~D. Moore, ``{Electroweak Bubble Wall Speed Limit},''
  \href{http://dx.doi.org/10.1088/1475-7516/2017/05/025}{{\em JCAP} {\bfseries
  05} (2017) 025}, \href{http://arxiv.org/abs/1703.08215}{{\ttfamily
  arXiv:1703.08215 [hep-ph]}}.

\bibitem{Gouttenoire:2021kjv}
Y.~Gouttenoire, R.~Jinno, and F.~Sala, ``{Friction pressure on relativistic
  bubble walls},'' \href{http://dx.doi.org/10.1007/JHEP05(2022)004}{{\em JHEP}
  {\bfseries 05} (2022) 004}, \href{http://arxiv.org/abs/2112.07686}{{\ttfamily
  arXiv:2112.07686 [hep-ph]}}.

\bibitem{Enqvist:1991xw}
K.~Enqvist, J.~Ignatius, K.~Kajantie, and K.~Rummukainen, ``{Nucleation and
  bubble growth in a first order cosmological electroweak phase transition},''
  \href{http://dx.doi.org/10.1103/PhysRevD.45.3415}{{\em Phys. Rev. D}
  {\bfseries 45} (1992) 3415--3428}.

\bibitem{Baldes:2023rqv}
I.~Baldes and M.~O. Olea-Romacho, ``{Primordial black holes as dark matter:
  Interferometric tests of phase transition origin},''
  \href{http://arxiv.org/abs/2307.11639}{{\ttfamily arXiv:2307.11639
  [hep-ph]}}.

\bibitem{Baldes:2023boh}
I.~Baldes, M.~Dichtl, Y.~Gouttenoire, and F.~Sala, ``{Particle Shells from
  Relativistic Bubble Walls},''
  \href{http://arxiv.org/abs/231X.XXXXX}{{\ttfamily arXiv:231X.XXXXX
  [hep-ph]}}.

\bibitem{Bruggisser:2018mus}
S.~Bruggisser, B.~Von~Harling, O.~Matsedonskyi, and G.~Servant, ``{Baryon
  Asymmetry from a Composite Higgs Boson},''
  \href{http://dx.doi.org/10.1103/PhysRevLett.121.131801}{{\em Phys. Rev.
  Lett.} {\bfseries 121} no.~13, (2018) 131801},
  \href{http://arxiv.org/abs/1803.08546}{{\ttfamily arXiv:1803.08546
  [hep-ph]}}.

\bibitem{Bruggisser:2022rdm}
S.~Bruggisser, B.~von Harling, O.~Matsedonskyi, and G.~Servant, ``{Status of
  electroweak baryogenesis in minimal composite Higgs},''
  \href{http://dx.doi.org/10.1007/JHEP08(2023)012}{{\em JHEP} {\bfseries 08}
  (2023) 012}, \href{http://arxiv.org/abs/2212.11953}{{\ttfamily
  arXiv:2212.11953 [hep-ph]}}.

\bibitem{Kuzmin:1985mm}
V.~A. Kuzmin, V.~A. Rubakov, and M.~E. Shaposhnikov, ``{On the Anomalous
  Electroweak Baryon Number Nonconservation in the Early Universe},''
  \href{http://dx.doi.org/10.1016/0370-2693(85)91028-7}{{\em Phys. Lett. B}
  {\bfseries 155} (1985) 36}.

\bibitem{Khlebnikov:1988sr}
S.~Y. Khlebnikov and M.~E. Shaposhnikov, ``{The Statistical Theory of Anomalous
  Fermion Number Nonconservation},''
  \href{http://dx.doi.org/10.1016/0550-3213(88)90133-2}{{\em Nucl. Phys. B}
  {\bfseries 308} (1988) 885--912}.

\bibitem{Cutting:2020nla}
D.~Cutting, E.~G. Escartin, M.~Hindmarsh, and D.~J. Weir, ``{Gravitational
  waves from vacuum first order phase transitions II: from thin to thick
  walls},'' \href{http://dx.doi.org/10.1103/PhysRevD.103.023531}{{\em Phys.
  Rev. D} {\bfseries 103} no.~2, (2021) 023531},
  \href{http://arxiv.org/abs/2005.13537}{{\ttfamily arXiv:2005.13537
  [astro-ph.CO]}}.

\bibitem{Durrer:2003ja}
R.~Durrer and C.~Caprini, ``{Primordial magnetic fields and causality},''
  \href{http://dx.doi.org/10.1088/1475-7516/2003/11/010}{{\em JCAP} {\bfseries
  11} (2003) 010}, \href{http://arxiv.org/abs/astro-ph/0305059}{{\ttfamily
  arXiv:astro-ph/0305059}}.

\bibitem{Caprini:2015zlo}
C.~Caprini {\em et~al.}, ``{Science with the space-based interferometer eLISA.
  II: Gravitational waves from cosmological phase transitions},''
  \href{http://dx.doi.org/10.1088/1475-7516/2016/04/001}{{\em JCAP} {\bfseries
  04} (2016) 001}, \href{http://arxiv.org/abs/1512.06239}{{\ttfamily
  arXiv:1512.06239 [astro-ph.CO]}}.

\bibitem{Caprini:2019egz}
C.~Caprini {\em et~al.}, ``{Detecting gravitational waves from cosmological
  phase transitions with LISA: an update},''
  \href{http://dx.doi.org/10.1088/1475-7516/2020/03/024}{{\em JCAP} {\bfseries
  03} (2020) 024}, \href{http://arxiv.org/abs/1910.13125}{{\ttfamily
  arXiv:1910.13125 [astro-ph.CO]}}.

\bibitem{Cutting:2019zws}
D.~Cutting, M.~Hindmarsh, and D.~J. Weir, ``{Vorticity, kinetic energy, and
  suppressed gravitational wave production in strong first order phase
  transitions},'' \href{http://dx.doi.org/10.1103/PhysRevLett.125.021302}{{\em
  Phys. Rev. Lett.} {\bfseries 125} no.~2, (2020) 021302},
  \href{http://arxiv.org/abs/1906.00480}{{\ttfamily arXiv:1906.00480
  [hep-ph]}}.

\bibitem{Jinno:2019jhi}
R.~Jinno, H.~Seong, M.~Takimoto, and C.~M. Um, ``{Gravitational waves from
  first-order phase transitions: Ultra-supercooled transitions and the fate of
  relativistic shocks},''
  \href{http://dx.doi.org/10.1088/1475-7516/2019/10/033}{{\em JCAP} {\bfseries
  10} (2019) 033}, \href{http://arxiv.org/abs/1905.00899}{{\ttfamily
  arXiv:1905.00899 [astro-ph.CO]}}.

\bibitem{Konstandin:2017sat}
T.~Konstandin, ``{Gravitational radiation from a bulk flow model},''
  \href{http://dx.doi.org/10.1088/1475-7516/2018/03/047}{{\em JCAP} {\bfseries
  03} (2018) 047}, \href{http://arxiv.org/abs/1712.06869}{{\ttfamily
  arXiv:1712.06869 [astro-ph.CO]}}.

\bibitem{Hild:2010id}
S.~Hild {\em et~al.}, ``{Sensitivity Studies for Third-Generation Gravitational
  Wave Observatories},''
  \href{http://dx.doi.org/10.1088/0264-9381/28/9/094013}{{\em Class. Quant.
  Grav.} {\bfseries 28} (2011) 094013},
  \href{http://arxiv.org/abs/1012.0908}{{\ttfamily arXiv:1012.0908 [gr-qc]}}.

\bibitem{Thrane:2013oya}
E.~Thrane and J.~D. Romano, ``{Sensitivity curves for searches for
  gravitational-wave backgrounds},''
  \href{http://dx.doi.org/10.1103/PhysRevD.88.124032}{{\em Phys. Rev. D}
  {\bfseries 88} no.~12, (2013) 124032},
  \href{http://arxiv.org/abs/1310.5300}{{\ttfamily arXiv:1310.5300
  [astro-ph.IM]}}.

\bibitem{Caprini:2019pxz}
C.~Caprini, D.~G. Figueroa, R.~Flauger, G.~Nardini, M.~Peloso, M.~Pieroni,
  A.~Ricciardone, and G.~Tasinato, ``{Reconstructing the spectral shape of a
  stochastic gravitational wave background with LISA},''
  \href{http://dx.doi.org/10.1088/1475-7516/2019/11/017}{{\em JCAP} {\bfseries
  11} (2019) 017}, \href{http://arxiv.org/abs/1906.09244}{{\ttfamily
  arXiv:1906.09244 [astro-ph.CO]}}.

\bibitem{Kawamura:2020pcg}
S.~Kawamura {\em et~al.}, ``{Current status of space gravitational wave antenna
  DECIGO and B-DECIGO},'' \href{http://dx.doi.org/10.1093/ptep/ptab019}{{\em
  PTEP} {\bfseries 2021} no.~5, (2021) 05A105},
  \href{http://arxiv.org/abs/2006.13545}{{\ttfamily arXiv:2006.13545 [gr-qc]}}.

\bibitem{KAGRA:2021kbb}
{\bfseries KAGRA, Virgo, LIGO Scientific} Collaboration, R.~Abbott {\em
  et~al.}, ``{Upper limits on the isotropic gravitational-wave background from
  Advanced LIGO and Advanced Virgo\textquoteright{}s third observing run},''
  \href{http://dx.doi.org/10.1103/PhysRevD.104.022004}{{\em Phys. Rev. D}
  {\bfseries 104} no.~2, (2021) 022004},
  \href{http://arxiv.org/abs/2101.12130}{{\ttfamily arXiv:2101.12130 [gr-qc]}}.

\bibitem{Branchesi:2023mws}
M.~Branchesi {\em et~al.}, ``{Science with the Einstein Telescope: a comparison
  of different designs},''
  \href{http://dx.doi.org/10.1088/1475-7516/2023/07/068}{{\em JCAP} {\bfseries
  07} (2023) 068}, \href{http://arxiv.org/abs/2303.15923}{{\ttfamily
  arXiv:2303.15923 [gr-qc]}}.

\bibitem{Rosado:2011kv}
P.~A. Rosado, ``{Gravitational wave background from binary systems},''
  \href{http://dx.doi.org/10.1103/PhysRevD.84.084004}{{\em Phys. Rev. D}
  {\bfseries 84} (2011) 084004},
  \href{http://arxiv.org/abs/1106.5795}{{\ttfamily arXiv:1106.5795 [gr-qc]}}.

\bibitem{Robson:2018ifk}
T.~Robson, N.~J. Cornish, and C.~Liu, ``{The construction and use of LISA
  sensitivity curves},'' \href{http://dx.doi.org/10.1088/1361-6382/ab1101}{{\em
  Class. Quant. Grav.} {\bfseries 36} no.~10, (2019) 105011},
  \href{http://arxiv.org/abs/1803.01944}{{\ttfamily arXiv:1803.01944
  [astro-ph.HE]}}.

\bibitem{Farmer:2003pa}
A.~J. Farmer and E.~S. Phinney, ``{The gravitational wave background from
  cosmological compact binaries},''
  \href{http://dx.doi.org/10.1111/j.1365-2966.2003.07176.x}{{\em Mon. Not. Roy.
  Astron. Soc.} {\bfseries 346} (2003) 1197},
  \href{http://arxiv.org/abs/astro-ph/0304393}{{\ttfamily
  arXiv:astro-ph/0304393}}.

\bibitem{Aggarwal:2018mgp}
K.~Aggarwal {\em et~al.}, ``{The NANOGrav 11-Year Data Set: Limits on
  Gravitational Waves from Individual Supermassive Black Hole Binaries},''
  \href{http://dx.doi.org/10.3847/1538-4357/ab2236}{{\em Astrophys. J.}
  {\bfseries 880} (2019) 2}, \href{http://arxiv.org/abs/1812.11585}{{\ttfamily
  arXiv:1812.11585 [astro-ph.GA]}}.

\bibitem{NANOGrav:2020bcs}
{\bfseries NANOGrav} Collaboration, Z.~Arzoumanian {\em et~al.}, ``{The
  NANOGrav 12.5 yr Data Set: Search for an Isotropic Stochastic
  Gravitational-wave Background},''
  \href{http://dx.doi.org/10.3847/2041-8213/abd401}{{\em Astrophys. J. Lett.}
  {\bfseries 905} no.~2, (2020) L34},
  \href{http://arxiv.org/abs/2009.04496}{{\ttfamily arXiv:2009.04496
  [astro-ph.HE]}}.

\bibitem{Goncharov:2021oub}
B.~Goncharov {\em et~al.}, ``{On the Evidence for a Common-spectrum Process in
  the Search for the Nanohertz Gravitational-wave Background with the Parkes
  Pulsar Timing Array},''
  \href{http://dx.doi.org/10.3847/2041-8213/ac17f4}{{\em Astrophys. J. Lett.}
  {\bfseries 917} no.~2, (2021) L19},
  \href{http://arxiv.org/abs/2107.12112}{{\ttfamily arXiv:2107.12112
  [astro-ph.HE]}}.

\bibitem{EPTA:2021crs}
{\bfseries EPTA} Collaboration, S.~Chen {\em et~al.}, ``{Common-red-signal
  analysis with 24-yr high-precision timing of the European Pulsar Timing
  Array: inferences in the stochastic gravitational-wave background search},''
  \href{http://dx.doi.org/10.1093/mnras/stab2833}{{\em Mon. Not. Roy. Astron.
  Soc.} {\bfseries 508} no.~4, (2021) 4970--4993},
  \href{http://arxiv.org/abs/2110.13184}{{\ttfamily arXiv:2110.13184
  [astro-ph.HE]}}.

\bibitem{Antoniadis:2022pcn}
J.~Antoniadis {\em et~al.}, ``{The International Pulsar Timing Array second
  data release: Search for an isotropic gravitational wave background},''
  \href{http://dx.doi.org/10.1093/mnras/stab3418}{{\em Mon. Not. Roy. Astron.
  Soc.} {\bfseries 510} no.~4, (2022) 4873--4887},
  \href{http://arxiv.org/abs/2201.03980}{{\ttfamily arXiv:2201.03980
  [astro-ph.HE]}}.

\bibitem{NANOGrav:2021flc}
{\bfseries NANOGrav} Collaboration, Z.~Arzoumanian {\em et~al.}, ``{Searching
  for Gravitational Waves from Cosmological Phase Transitions with the NANOGrav
  12.5-Year Dataset},''
  \href{http://dx.doi.org/10.1103/PhysRevLett.127.251302}{{\em Phys. Rev.
  Lett.} {\bfseries 127} no.~25, (2021) 251302},
  \href{http://arxiv.org/abs/2104.13930}{{\ttfamily arXiv:2104.13930
  [astro-ph.CO]}}.

\bibitem{Bringmann:2023opz}
T.~Bringmann, P.~F. Depta, T.~Konstandin, K.~Schmidt-Hoberg, and C.~Tasillo,
  ``{Does NANOGrav observe a dark sector phase transition?},''
  \href{http://arxiv.org/abs/2306.09411}{{\ttfamily arXiv:2306.09411
  [astro-ph.CO]}}.

\bibitem{1960JMP.....1..429C}
R.~E. {Cutkosky}, ``{Singularities and Discontinuities of Feynman
  Amplitudes},'' \href{http://dx.doi.org/10.1063/1.1703676}{{\em Journal of
  Mathematical Physics} {\bfseries 1} no.~5, (Sept., 1960) 429--433}.

\bibitem{Super-Kamiokande:2014otb}
{\bfseries Super-Kamiokande} Collaboration, K.~Abe {\em et~al.}, ``{Search for
  proton decay via $p\to\nu K^+$ using 260 kiloton\textperiodcentered{}year
  data of Super-Kamiokande},''
  \href{http://dx.doi.org/10.1103/PhysRevD.90.072005}{{\em Phys. Rev. D}
  {\bfseries 90} no.~7, (2014) 072005},
  \href{http://arxiv.org/abs/1408.1195}{{\ttfamily arXiv:1408.1195 [hep-ex]}}.

\bibitem{ALEPH:2005ab}
{\bfseries ALEPH, DELPHI, L3, OPAL, SLD, LEP Electroweak Working Group, SLD
  Electroweak Group, SLD Heavy Flavour Group} Collaboration, S.~Schael {\em
  et~al.}, ``{Precision electroweak measurements on the $Z$ resonance},''
  \href{http://dx.doi.org/10.1016/j.physrep.2005.12.006}{{\em Phys. Rept.}
  {\bfseries 427} (2006) 257--454},
  \href{http://arxiv.org/abs/hep-ex/0509008}{{\ttfamily arXiv:hep-ex/0509008}}.

\bibitem{ALEPH:2002nmz}
{\bfseries ALEPH} Collaboration, A.~Heister {\em et~al.}, ``{Search for
  supersymmetric particles with R parity violating decays in $e^{+} e^{-}$
  collisions at $\sqrt{s}$ up to 209-GeV},''
  \href{http://dx.doi.org/10.1140/epjc/s2003-01311-5}{{\em Eur. Phys. J. C}
  {\bfseries 31} (2003) 1--16},
  \href{http://arxiv.org/abs/hep-ex/0210014}{{\ttfamily arXiv:hep-ex/0210014}}.

\bibitem{CDF:2013yum}
{\bfseries CDF} Collaboration, T.~Aaltonen {\em et~al.}, ``{Search for Pair
  Production of Strongly Interacting Particles Decaying to Pairs of Jets in $p
  \bar{p}$ Collisions at $\sqrt{s} =$ 1.96 TeV},''
  \href{http://dx.doi.org/10.1103/PhysRevLett.111.031802}{{\em Phys. Rev.
  Lett.} {\bfseries 111} no.~3, (2013) 031802},
  \href{http://arxiv.org/abs/1303.2699}{{\ttfamily arXiv:1303.2699 [hep-ex]}}.

\bibitem{ATLAS:2017jnp}
{\bfseries ATLAS} Collaboration, M.~Aaboud {\em et~al.}, ``{A search for
  pair-produced resonances in four-jet final states at $\sqrt{s} =$ 13 TeV with
  the ATLAS detector},''
  \href{http://dx.doi.org/10.1140/epjc/s10052-018-5693-4}{{\em Eur. Phys. J. C}
  {\bfseries 78} no.~3, (2018) 250},
  \href{http://arxiv.org/abs/1710.07171}{{\ttfamily arXiv:1710.07171
  [hep-ex]}}.

\bibitem{CMS:2022usq}
{\bfseries CMS} Collaboration, ``{Search for resonant and nonresonant
  production of pairs of dijet resonances in proton-proton collisions at
  $\sqrt{s}$ = 13 TeV},'' \href{http://arxiv.org/abs/2206.09997}{{\ttfamily
  arXiv:2206.09997 [hep-ex]}}.

\bibitem{Abel:2020pzs}
C.~Abel {\em et~al.}, ``{Measurement of the Permanent Electric Dipole Moment of
  the Neutron},'' \href{http://dx.doi.org/10.1103/PhysRevLett.124.081803}{{\em
  Phys. Rev. Lett.} {\bfseries 124} no.~8, (2020) 081803},
  \href{http://arxiv.org/abs/2001.11966}{{\ttfamily arXiv:2001.11966
  [hep-ex]}}.

\bibitem{Bensalem:2021qtj}
W.~Bensalem and D.~Stolarski, ``{Flavor and CP violation from a QCD-like hidden
  sector},'' \href{http://dx.doi.org/10.1007/JHEP02(2022)011}{{\em JHEP}
  {\bfseries 02} (2022) 011}, \href{http://arxiv.org/abs/2111.05515}{{\ttfamily
  arXiv:2111.05515 [hep-ph]}}.

\bibitem{Barr:1990vd}
S.~M. Barr and A.~Zee, ``{Electric Dipole Moment of the Electron and of the
  Neutron},'' \href{http://dx.doi.org/10.1103/PhysRevLett.65.21}{{\em Phys.
  Rev. Lett.} {\bfseries 65} (1990) 21--24}. [Erratum: Phys.Rev.Lett. 65, 2920
  (1990)].

\bibitem{Sala:2013osa}
F.~Sala, ``{A bound on the charm chromo-EDM and its implications},''
  \href{http://dx.doi.org/10.1007/JHEP03(2014)061}{{\em JHEP} {\bfseries 03}
  (2014) 061}, \href{http://arxiv.org/abs/1312.2589}{{\ttfamily arXiv:1312.2589
  [hep-ph]}}.

\bibitem{Giudice:2011ak}
G.~F. Giudice, B.~Gripaios, and R.~Sundrum, ``{Flavourful Production at Hadron
  Colliders},'' \href{http://dx.doi.org/10.1007/JHEP08(2011)055}{{\em JHEP}
  {\bfseries 08} (2011) 055}, \href{http://arxiv.org/abs/1105.3161}{{\ttfamily
  arXiv:1105.3161 [hep-ph]}}.

\bibitem{Bona:2022zhn}
M.~Bona {\em et~al.}, ``{Unitarity Triangle global fits beyond the Standard
  Model: UTfit 2021 NP update},''
  \href{http://dx.doi.org/10.22323/1.398.0500}{{\em PoS} {\bfseries
  EPS-HEP2021} (2022) 500}.

\bibitem{Fukugita:1986hr}
M.~Fukugita and T.~Yanagida, ``{Baryogenesis Without Grand Unification},''
  \href{http://dx.doi.org/10.1016/0370-2693(86)91126-3}{{\em Phys. Lett. B}
  {\bfseries 174} (1986) 45--47}.

\bibitem{Mohapatra:1986aw}
R.~N. Mohapatra, ``{Mechanism for Understanding Small Neutrino Mass in
  Superstring Theories},''
  \href{http://dx.doi.org/10.1103/PhysRevLett.56.561}{{\em Phys. Rev. Lett.}
  {\bfseries 56} (1986) 561--563}.

\bibitem{Mohapatra:1986bd}
R.~N. Mohapatra and J.~W.~F. Valle, ``{Neutrino Mass and Baryon Number
  Nonconservation in Superstring Models},''
  \href{http://dx.doi.org/10.1103/PhysRevD.34.1642}{{\em Phys. Rev. D}
  {\bfseries 34} (1986) 1642}.

\bibitem{Weinberg:1979sa}
S.~Weinberg, ``{Baryon and Lepton Nonconserving Processes},''
  \href{http://dx.doi.org/10.1103/PhysRevLett.43.1566}{{\em Phys. Rev. Lett.}
  {\bfseries 43} (1979) 1566--1570}.

\bibitem{Chacko:2020zze}
Z.~Chacko, P.~J. Fox, R.~Harnik, and Z.~Liu, ``{Neutrino Masses from Low Scale
  Partial Compositeness},''
  \href{http://dx.doi.org/10.1007/JHEP03(2021)112}{{\em JHEP} {\bfseries 03}
  (2021) 112}, \href{http://arxiv.org/abs/2012.01443}{{\ttfamily
  arXiv:2012.01443 [hep-ph]}}.

\bibitem{Agashe:2018cuf}
K.~Agashe, P.~Du, M.~Ekhterachian, C.~S. Fong, S.~Hong, and L.~Vecchi,
  ``{Natural Seesaw and Leptogenesis from Hybrid of High-Scale Type I and
  TeV-Scale Inverse},'' \href{http://dx.doi.org/10.1007/JHEP04(2019)029}{{\em
  JHEP} {\bfseries 04} (2019) 029},
  \href{http://arxiv.org/abs/1812.08204}{{\ttfamily arXiv:1812.08204
  [hep-ph]}}.

\bibitem{Covi:1996wh}
L.~Covi, E.~Roulet, and F.~Vissani, ``{CP violating decays in leptogenesis
  scenarios},'' \href{http://dx.doi.org/10.1016/0370-2693(96)00817-9}{{\em
  Phys. Lett. B} {\bfseries 384} (1996) 169--174},
  \href{http://arxiv.org/abs/hep-ph/9605319}{{\ttfamily arXiv:hep-ph/9605319}}.

\bibitem{Deppisch:2010fr}
F.~F. Deppisch and A.~Pilaftsis, ``{Lepton Flavour Violation and theta(13) in
  Minimal Resonant Leptogenesis},''
  \href{http://dx.doi.org/10.1103/PhysRevD.83.076007}{{\em Phys. Rev. D}
  {\bfseries 83} (2011) 076007},
  \href{http://arxiv.org/abs/1012.1834}{{\ttfamily arXiv:1012.1834 [hep-ph]}}.

\bibitem{Blanchet:2009kk}
S.~Blanchet, T.~Hambye, and F.-X. Josse-Michaux, ``{Reconciling leptogenesis
  with observable mu ---\ensuremath{>} e gamma rates},''
  \href{http://dx.doi.org/10.1007/JHEP04(2010)023}{{\em JHEP} {\bfseries 04}
  (2010) 023}, \href{http://arxiv.org/abs/0912.3153}{{\ttfamily arXiv:0912.3153
  [hep-ph]}}.

\bibitem{AlAli:2021let}
H.~Al~Ali {\em et~al.}, ``{The muon Smasher\textquoteright{}s guide},''
  \href{http://dx.doi.org/10.1088/1361-6633/ac6678}{{\em Rept. Prog. Phys.}
  {\bfseries 85} no.~8, (2022) 084201},
  \href{http://arxiv.org/abs/2103.14043}{{\ttfamily arXiv:2103.14043
  [hep-ph]}}.

\bibitem{Aime:2022flm}
C.~Aime {\em et~al.}, ``{Muon Collider Physics Summary},''
  \href{http://arxiv.org/abs/2203.07256}{{\ttfamily arXiv:2203.07256
  [hep-ph]}}.

\bibitem{Li:2023tbx}
P.~Li, Z.~Liu, and K.-F. Lyu, ``{Heavy neutral leptons at muon colliders},''
  \href{http://dx.doi.org/10.1007/JHEP03(2023)231}{{\em JHEP} {\bfseries 03}
  (2023) 231}, \href{http://arxiv.org/abs/2301.07117}{{\ttfamily
  arXiv:2301.07117 [hep-ph]}}.

\bibitem{Dolan:2018qpy}
M.~J. Dolan, T.~P. Dutka, and R.~R. Volkas, ``{Dirac-Phase Thermal Leptogenesis
  in the extended Type-I Seesaw Model},''
  \href{http://dx.doi.org/10.1088/1475-7516/2018/06/012}{{\em JCAP} {\bfseries
  06} (2018) 012}, \href{http://arxiv.org/abs/1802.08373}{{\ttfamily
  arXiv:1802.08373 [hep-ph]}}.

\end{thebibliography}\endgroup
\end{document}